# Bolide infrasound signal morphology and yield estimates: A case study of two events detected by a dense acoustic sensor network

**Trevor C. Wilson[1,2*], Elizabeth A. Silber[1], Thomas A. Colston[2], Brian R. Elbing[2], Thom R. Edwards[1]**

[1] Sandia National Laboratories, Albuquerque, NM
[2] Oklahoma State University, Stillwater, OK

*Corresponding author: trevor.wilson@okstate.edu

# Abstract

Two bolides (2 June 2016 and 4 April 2019) were detected at multiple regional infrasound stations with many of the locations receiving multiple detections. Analysis of the received signals was used to estimate the yield, location and trajectory, and the type of shock that produced the received signal. The results from the infrasound analysis were compared with ground truth information that was collected through other sensing modalities. This multi-modal framework offers an expanded perspective on the processes governing bolide shock generation and propagation. The majority of signal features showed reasonable agreement between the infrasound-based interpretation and the other observational modalities, though the yield estimate from the 2019 bolide was significantly lower using the infrasound detections. There was also evidence suggesting that one of the detections was from a cylindrical shock that was initially propagating upward, which is unusual though not impossible.

# 1 Introduction

Characterization of meteoroids and asteroids entering the Earth's atmosphere is of immense scientific significance, from studying properties of small bodies in the solar system to evaluating impact threat (Chyba et al. 1993; Chapman 2008). Approximately $10^5$ tons/year (Plane 2012) of extraterrestrial material enters the Earth's atmosphere at speeds between 11.2 and 72.5 km/s (Ceplecha et al. 1998). Most of this material is comprised of small dust size particles with only a relatively small portion being centimeter-sized or greater, collectively called meteoroids and asteroids. As a meteoroid encounters dense regions of the Earth's atmosphere, it undergoes a number of physical and chemical processes, including heating, sputtering, and ablation. These processes result in luminous phenomena, generally referred to as a meteor. Very bright meteors are referred to as fireballs and bolides (Ceplecha et al. 1998).

Whether a meteoroid generates a shockwave upon entering denser regions of the atmosphere depends on the size, velocity, and Knudsen number (Silber et al. 2018). Knudsen number is the ratio of the mean free path of air molecules and the characteristic length scale of the object (see Silber et al. (2018). Objects in the size ranging from meters to tens of meters might form impact craters (Brown et al. 2008; Pichon et al. 2008; Kenkmann et al. 2009) and pose a serious threat to people and infrastructure, as exemplified by the Chelyabinsk bolide event in 2013 (Brown et al. 2013; Popova et al. 2013). Meteoroids, owing to their hypersonic velocities, intense rate of ablation, and possible fragmentation events, generate shockwaves (ballistic shock or cylindrical line source) as they enter the Earth's atmosphere (ReVelle 1976; Ceplecha et al. 1998; Silber & Brown 2014). For fragmentation events, the shockwave could be quasi-spherical or point source, depending on the type of fragmentation (continuous fragmentation, discrete fragmentation episodes, or a terminal burst). It is possible for some or all types of fragmentation to be present during a meteoroid's entry (Trigo-Rodríguez et al. 2021). Ablation enhances meteor generated shockwaves as well as shifts their onset to higher altitudes (Moreno-Ibáñez et al. 2018; Silber et al. 2018).

A byproduct of shockwave decay is infrasound, a low frequency ($f$) acoustic wave below the threshold of human hearing that can be detected by microbarometers. Typically, these sensors are installed on the ground (Fehr, 1967; Cook & Bedard Jr, 1971; Brachet et al. 2009), but in recent years they have also been deployed on floating platforms (e.g., Bowman et al. 2018; Brissaud et al. 2021; Silber et al. 2023; Silber et al. 2024). To obtain the infrasound wave direction and angle of arrival, an array of three or more microbarometers is typically used. There are many sources of infrasound, both anthropogenic (re-entry, explosions, rockets (ReVelle & Edwards 2007; Pilger et al. 2021; Silber et al. 2023)) and natural (bolides, lightning, volcanoes (Assink et al. 2008; Matoza et al. 2019; Pilger et al. 2020; Ott et al. 2021; Fernando et al. 2024)). Since infrasound can propagate over long distances with little attenuation, it is used as one of the four sensing modalities of the International Monitoring System (IMS) towards the Preparatory Commission of the Comprehensive Nuclear-Test-Ban Treaty compliance verification (Brachet et al. 2009; Christie & Campus 2010).

Infrasound monitoring is a valuable tool for detecting and geolocating bolides as well as inferring their yield (ReVelle 1997; Silber et al. 2009; Brown et al. 2013; Silber & Brown 2019; Pilger et al. 2020). Energy deposition estimates can provide useful calibrations for the IMS (Ens et al. 2012; Gi & Brown, 2017). Ens et al. (2012) reported that bolides with energies above 20 kt of trinitrotoluene (TNT) equivalent (1 kt of TNT = $4.184 \times 10^{12}$ joules) can be detected globally, while 1 kt events have a detection threshold around 7,000 km. One notable example of an infrasound-based bolide characterization is the detection of a daylight impact over Indonesia in 2009 (Silber et al. 2011). This energetic event was detected by 17 infrasound stations globally, with the most distant station situated 17,500 km from the event. Silber et al. (2011) determined the airburst altitude and energy deposition through infrasound records alone. At the time, no other observational data besides casual witness reports existed.

However, it is rare to detect a bolide at numerous stations because large energetic events are statistically less probable than weaker ones (Boslough et al. 2015; Harris & Chodas 2021; Silber et al. 2018). Excluding the historical bolides detected by infrasound in the 1960s and 1970s (Silber et al. 2009), there are only a handful of documented energetic bolide events detected globally, including the Chelyabinsk bolide (440 kt of TNT) (Brown et al. 2013) and the Bering Sea bolide (43 kt of TNT) (Arrowsmith et al. 2021). Since the inception of the IMS, there have been five documented bolide events, including that over Indonesia (Silber et al. 2011), with a yield greater than 20 kt. Other documented bolide events are much less energetic, and thus less likely to be detectable at longer ranges, diminishing the opportunity to study signals recorded by multiple infrasound stations. Therefore, for advancing infrasound studies towards better characterization of bolides, regardless of their energetics, it would be advantageous to utilize regional infrasound assets that have a relatively dense geographical coverage. For instance, a deployment of dense networks has proven highly effective during the Hayabusa 2 and Origins, Spectral Interpretation, Resource Identification, and Security-Regolith Explorer (OSIRIS-REx) sample return capsule re-entries (e.g., Sansom et al., 2022; Silber et al. 2024), demonstrating their potential to significantly expand the detection of both natural and artificial hypersonic sources (e.g., bolides, space debris, space missions). There are numerous geophysical sensing networks operating in the U.S., some of which also feature infrasound stations (arrays and single sensor).

This study presents detailed analyses of two bolide events that occurred over the U.S. (2 June 2016 and 4 April 2019), both of which were detected by multiple regional infrasound stations. Infrasound data from several networks were leveraged to explore the potential and efficacy of acoustic sensing (array and single sensor stations) for accurately evaluating and constraining bolide entry parameters. We further utilize ground truth and observations obtained from other sensing modalities. This paper is organized as follows: Section 2 presents approaches to detect and observe bolide events as well as ground truth for the two case studies; Section 3 describes the infrasound network and analysis methods; the results are presented and discussed in Section 4 (Section 4.1 – June 2016 bolide; Section 4.2 April 2019 bolide); and conclusions drawn in Section 5.

# 2 Observations of Bolides and Two Case Studies

## *2.1 A Brief Overview of Observational Modalities*

Infrasound sensing of meteoroid impacts can aid in constraining geolocation and estimating yield. Infrasound can be invaluable when other sensing modalities are unavailable (e.g., Hicks et al. 2023) or only scarcely available (see Silber 2024a). Nevertheless, while infrasound is a robust detection method, its limitations, such as dependence on atmospheric conditions for signal propagation and its detection range being influenced by the source signal strength, should be acknowledged (de Groot-Hedlin et al. 2009; Drob 2019; Drob et al. 2010). For this reason, it is preferrable to have ground truth (GT) information from other observational methods, some of which will be briefly mentioned here.

Notable observational approaches used for the study of meteors include eyewitness reports, optical (photograph/video or spectral methods), and space-based instruments. While eyewitness reports provide valuable context (Ahn 2003; Littmann & Suomela 2014), they are subjective with variable accuracy and reliability (Moser 2017). Optical sensing is the most common instrument-based method for detecting meteors (e.g., Madiedo & Trigo-Rodriguez 2008; Koten et al. 2011; Gritsevich 2008; Gritsevich & Stulov 2008; Gritsevich 2009; Sansom et al. 2019), ranging from amateur photographs/videos to specialized all-sky cameras (e.g., Devillepoix et al. 2020). Under adequate conditions (i.e., clear skies at night), optical observations can provide trajectory, velocity, and brightness (McCrosky & Boeschenstein Jr 1965; Gritsevich & Stulov 2006) of a meteoroid.

Satellite-based observations are an important source of fireball data due to vast sections of the atmosphere being continuously monitored (Nemtchinov et al. 1997). Thus, within the surveyed region there is a probability of detection if the event is sufficiently bright. U.S. government (USG) sensors provide global event information such as time, location, radiated energy, estimated impact energy, and in some cases peak brightness altitude and object velocity (Nemtchinov et al. 1997; Peña-Asensio et al. 2022). Another notable example is the Geostationary Lightning Mapper (GLM) (Jenniskens et al. 2018), deployed onboard the National Oceanic and Atmospheric Administration (NOAA) Geostationary Operational Environmental Satellites (GOES) 16 and 17. Its original purpose is to observe and record lightning events (Rudlosky et al. 2019) but has recently demonstrated the capability to detect bright fireballs (Jenniskens et al. 2018; Ott et al. 2021). An algorithm was developed utilizing GLM data to identify fireball signatures by exploiting differences between the ground tracks and light curves of lightning and bolides (Jenniskens et al., 2018). The main GLM sensor limitations are their field of view (confined to North and South America) and that stereo observations require an overlap of coverage areas.

In recent years, several publicly accessible databases have compiled numerous meteor reports. The International Meteor Organization (IMO) and the American Meteor Society (AMS) fireball catalogues are examples of witness report databases, where anyone can report their visual or auditory observations (descriptions, photographs, and videos). With enough reports, rough estimates of the altitude, location, and trajectory are sometimes possible. However, these reports are best used in combination with other, more robust sources.

Satellite-derived databases include the Center for Near-Earth Object Studies (CNEOS) USG sensor fireball catalogue (*https://cneos.jpl.nasa.gov/fireballs/)* and the National Aeronautics and Space Administration's (NASA) Jet Propulsion Laboratory (JPL) hosted GLM-derived fireball database (*https://neo-bolide.ndc.nasa.gov*). The CNEOS fireball catalogue includes time, location, total radiated energy, and calculated total impact energy for all observed fireballs. As of recently, light curves are also available and ~30% of events list the altitude of peak brightness and velocity vector. The GLM-derived fireball database includes many previously unreported or unidentified events that were found using the fireball detection algorithm (Smith et al. 2021).

A 'light curve' is a plot of the observed brightness as a function of time of an object during the luminous portion of its flight through the atmosphere (Jacchia 1955). Light curve features can provide clues about distinct events during entry (strong ablation, continuous fragmentation, and gross fragmentation) as well as help in constraining other parameters of interest such as energy and luminous flight duration (Hawkes et al. 2001; Brosch et al. 2004; Subasinghe et al. 2016). The light curve signatures can indicate single or multiple fragmentation events (flares or sudden spikes in brightness), continuous fragmentation (smaller, recurrent brightness peaks), or non-fragmentation (steady rise in intensity during ablation as more mass is lost as it enters denser regions of the atmosphere). These light curves can provide critical ground truth in meteoroid characterization and analysis (e.g., Drolshagen et al. 2021; Gritsevich & Koschny 2011), especially when comparing to infrasound observations (e.g., Silber 2024a,b).

In this study, infrasound is the primary data source with several of the other methods providing context and ground truth for the bolide events investigated. Two events over the continental USA were selected for this study due to their proximity to a number of infrasound stations: a bolide over central Arizona in 2016 (Wilson et al. 2023b; Jenniskens et al. 2020; Palotai et al. 2019) and a bolide over western Arkansas in 2019 (Wilson et al. 2023a).

## 2.2 June 2016 Bolide

A bolide, later named "Dishchii'bikoh," entered the Earth's atmosphere over central Arizona on 2 June 2016 at 10:56:26 UTC (03:56 AM local time). Herein, this bolide will be referred to as the "2016 event." This event had 422 eyewitness reports submitted to IMO due to it occurring in the early morning hours while the skies were still dark. Images[1] and video[2] show several visible flares, a dust trail that lingered until sunrise, and a likely north to south trajectory.

[1] https://fireball.imo.net/members/imo_photo/view_photo?photo_id=5501
[2] https://www.youtube.com/watch?v=5fOMycoEx0Y

The IMO reports were used to estimate that the meteoroid ground track had a length of 40 km with a heading of 172 degrees (nearly due south). The onset and termination of the luminous path were at altitudes of ~80 km and ~33 km, respectively. The detection is listed on the CNEOS database with a reported ground location at maximum intensity of 33.8 N, 110.9 W and an estimated impact energy of 0.49 kt. The peak brightness altitude was not made available. The light curve from the CNEOS database has been digitized in Figure 1. Based on the variations in brightness intensity, the light curve suggests strong ablation for the first 0.5 seconds followed by two fragmentation episodes with the second fragmentation being terminal. This is supported by the video record that shows distinct fragmentation episodes (see Figure 1 inset).

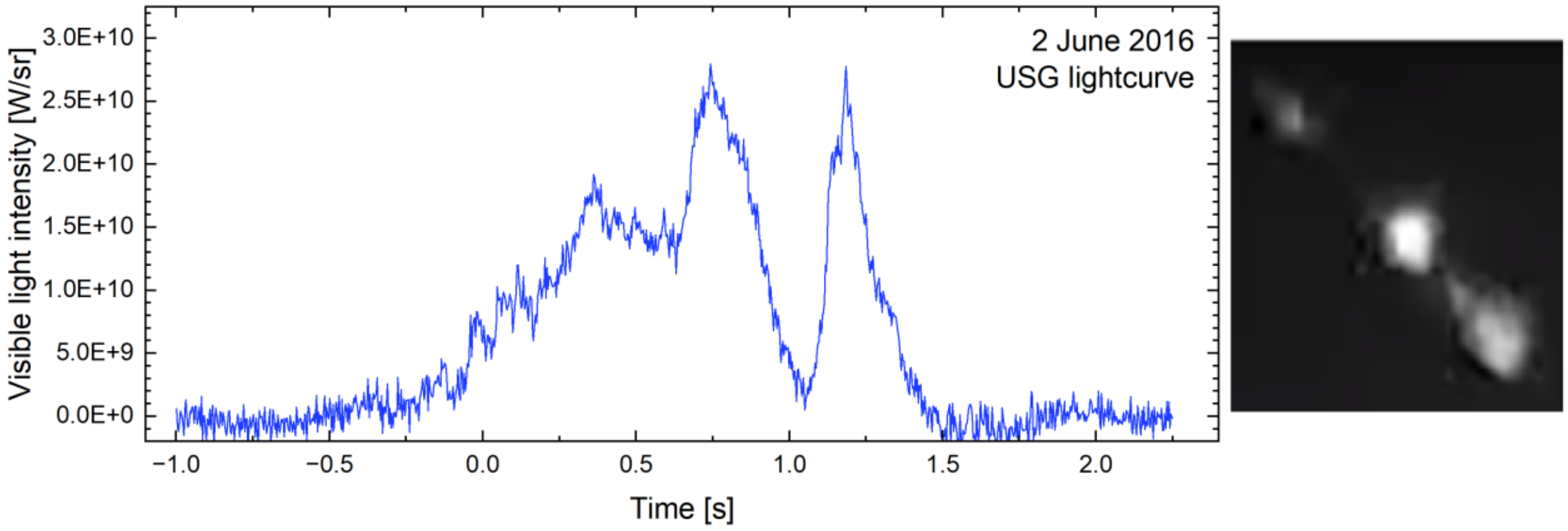

Figure 1. Digitized light curve from the CNEOS fireball database of the 2 June 2016 bolide using a 6000 K blackbody model. The inset image is from home security footage of the bolide showing what appears to be three distinct flares. Video/image credit: Justin D., reported sighting on IMO fireball database[3].

Two studies aimed to constrain the ground truth information of this event by leveraging photographic and video observations. The first utilized a combination of the SkySentinel's Spalding Allsky Camera Network (SACN) and a casual dash cam video (Palotai et al. 2019). They were able to estimate an onset location and altitude from the SACN network and the terminal point from the casual dash cam video. The terminal point has large uncertainty because it is solely based on the dash cam, which lacks proper calibration. Subsequently, Jenniskens et al. (2020) studied this event using a methodology involving the Lowell Observatory CAMS (LO-CAMS) low-light video camera network, the SACN (Palotai et al. 2019), a Multi-Spectral Radiometer, astrometric modeling, and the IMO reports to characterize the trajectory and timeline of the meteoroid flight. In addition, a 917 mg sample of one of the recovered fragments from this meteoroid was analyzed. These additional data further constrained the entry and origin of this meteoroid. Table 1 compares the reported characteristics of this meteoroid (i.e., Dishchii'bikoh) from these two studies.

[3] https://fireball.amsmeteors.org/members/imo_video/view_video?video_id=328

Table 1. Ground truth information for the 2016 event from previous studies and USG. Note that end location is the end of the record measured by the sensor network, not the terminal point. The latitude and longitude positions for USG were defined as 'end' as it is the only provided position from this source.

| Parameter | Palotai et al. (2019) | Jenniskens et al. (2020) | USG |
|---|---|---|---|
| **Entry latitude (°N)** | 34.555 | 34.647 | - |
| **Entry longitude (°W)** | 110.459 | 110.472 | - |
| **Luminous onset altitude (km)** | 100.2 ± 0.4 | 108.5 | - |
| **Entry velocity (km/s)** | 17.4 ± 0.3 | 16.6 ± 0.2 | - |
| **End latitude (°N)** | 33.924 | 34.391 | 33.8 |
| **End longitude (°W)** | 110.641 | 110.541 | 110.9 |
| **End altitude (km)** | 21.9 ± 0.6 | 75.06 | - |
| **Estimated energy (kt)** | 0.54 | 0.047 | 0.49 |
| **Diameter (m)** | 2 | 0.80 ± 0.2 | - |
| **Mass (metric tonnes)** | 14.8 ± 1.7 | 1.05 | - |
| **Density (kg $m^{-3}$)** | 2000–3000 | 3500 | - |
| **Flare event altitudes (km)** | - | 34, 29, 25 | - |
| **Initial heading (°)** | - | 192.5 | - |
| **Entry angle (°)** | - | 47.4 | - |

## *2.3 April 2019 Bolide*

On 4 April 2019 at 22:19:01 UTC (05:19 PM local time), a daylight bolide passed over northwestern Arkansas. This event will herein be referred to as the '2019 event'. The CNEOS database reports the location of peak brightness at 35.3 N, -93.9 W with an estimated total impact energy of 0.098 kt. However, the altitude was not reported. While eyewitness observations were limited, a smoke trail was photographed by a casual observer on the ground[4] and there were reports of an audible sonic boom[5]. The digitized CNEOS light curve in Figure 2 suggest that the meteoroid experienced two fragmentation episodes, which is corroborated by the smoke trail photograph

[4] https://fireball.imo.net/members/imo_photo/view_photo?photo_id=4227
[5] https://www.5newsonline.com/article/weather/meteor-causes-sonic-boom-thursday-evening/527-28950677-7285-474e-b7de-33bba0c93255

(Figure 2). The light curve also indicates that the first fragmentation episode was more energetic than the subsequent, which was likely a terminal fragmentation episode. This event was also reported in the GLM bolide database.

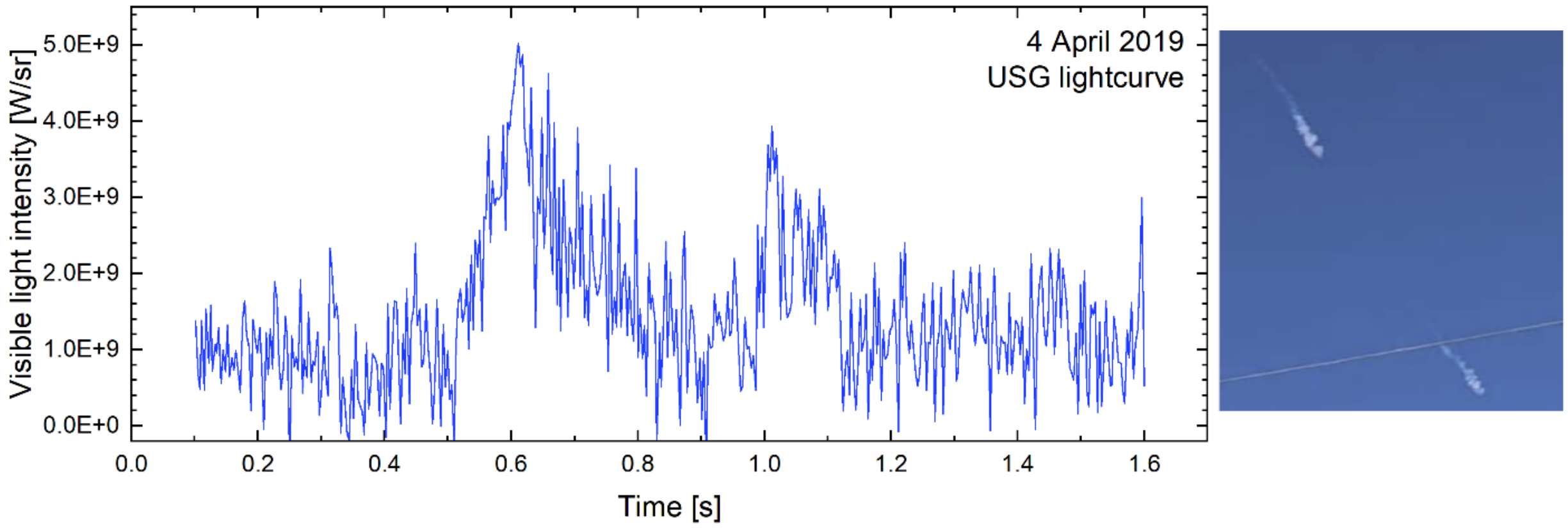

Figure 2. Light curve of the 4 April 2019 bolide showing light intensity versus time, digitized form CNEOS fireball database, using 6000 K blackbody model. Photo on the right, taken by Deb Vonnahme, shows the likely fragmentation episodes this meteoroid experienced during entry[4].

# 3 Methods

## 3.1 *Search for Signals at Infrasound Stations*

An estimated time of arrival technique was used to narrow the temporal search window (Silber & Brown 2014; Ruiz et al. 2013; Cobos et al. 2017, Silber 2024b) for both events. The signal travel times to potential infrasound stations with available data (up to 6250 km from the events) were estimated given the great circle distance between the USG-reported event location and the station divided by the celerity accounting for possible acoustic ducts (or waveguides). Arrival time windows for each station were computed for tropospheric (310–330 m/s), stratospheric (280-310 m/s), and thermospheric (180-280 m/s) waveguides (e.g., Negraru et al. 2010). The windows were extended for a subset of stations (U38B, T35B, and TUL3) during the 2019 event due to the relatively short distance between the event and the station combined with large uncertainty in the event location (i.e., >100 km difference between GLM and CNEOS). For clarity, stations with confirmed detections within the windows of interest are listed in Table 2. For the 2016 event and 2019 event there were three (O20A, KSCO, R11A) and four (S39B, T42B, Y45B, 143B) stations, respectively, searched without a confirmed detection.

Table 2. Infrasound stations investigated for detection of signals generated by the 2016 or the 2019 bolide events (stations without an observed detection are marked with an *). Array aperture is defined as the maximum distance between two nodes of an array.

| Station | Distance (km) | Latitude (° N) | Longitude (° E) | Sensor Model | Network | Sampling Rate (Hz) | # of Sensors | Aperture (m) |
|---|---|---|---|---|---|---|---|---|
| *June 2016 Bolide* | | | | | | | | |
| 121A | 322.39 | 32.532 | -107.785 | Hyperion 4000 | TA | 40 | 1 | N/A |
| 214A | 271.88 | 31.956 | -112.812 | Hyperion 4000 | TA | 40 | 1 | N/A |
| I57 | 514.68 | 33.599 | -116.459 | MB2000 / Chaparral 50 | IM | 20 | 8 | 1450 |
| I58 | 6256.51 | 28.203 | -177.384 | Chaparral 50 | IM | 20 | 4 | 1600 |
| I59 | 4698.12 | 19.586 | -155.899 | Chaparral 50 | IM | 20 | 4 | 1900 |
| IS1 | 602.39 | 37.223 | -116.061 | Hyperion 5000/3000 | SN | 200 | 4 | 55 |
| IS2 | 602.12 | 37.219 | -116.061 | Hyperion 5000/3000 | SN | 200 | 4 | 55 |
| IS3 | 602.53 | 37.222 | -116.064 | Hyperion 5000/3000 | SN | 200 | 4 | 55 |
| IS4 | 602.18 | 37.223 | -116.058 | Hyperion 5000/3000 | SN | 200 | 4 | 55 |
| IS7 | 600.82 | 37.203 | -116.058 | Hyperion 3000 | SN | 200 | 4 | 55 |
| KSCO* | 939.72 | 39.011 | -102.627 | Hyperion 4000 | TA | 40 | 1 | N/A |
| MR | 585.1 | 32.890 | -117.105 | Chaparral 50 | UH | 40 | 6 | 80 |
| O20A* | 743.08 | 40.135 | -108.241 | Hyperion 4000 | TA | 40 | 1 | N/A |
| R11A* | 658.12 | 38.349 | -115.585 | Hyperion 4000 | TA | 40 | 1 | N/A |
| TPFO | 514.17 | 33.606 | -116.454 | Hyperion 4000 | N4 | 40 | 1 | N/A |
| W18A | 181.29 | 35.118 | -109.736 | Hyperion 4000 | TA | 40 | 1 | N/A |
| Y22D | 368.31 | 34.074 | -106.921 | Hyperion 4000 | TA | 40 | 1 | N/A |
| *April 2019 Bolide* | | | | | | | | |
| 143B* | 492.5 | 32.703 | -91.404 | Hyperion 4000 | N4 | 40 | 1 | N/A |
| OSU1 | 211.2 | 36.134 | -97.082 | Chaparral 24 | OS | 1000 | 3 | 60 |
| R32B | 432.3 | 38.422 | -98.711 | Hyperion 4000 | N4 | 40 | 1 | N/A |
| S39B* | 206.1 | 37.691 | -93.323 | Hyperion 4000 | N4 | 40 | 1 | N/A |
| T35B | 179.7 | 36.916 | -96.512 | Hyperion 4000 | N4 | 40 | 1 | N/A |
| T42B* | 334.9 | 37.030 | -91.093 | Hyperion 4000 | N4 | 40 | 1 | N/A |
| TUL3 | 99.6 | 35.911 | -95.792 | Hyperion 4000 | N4 | 40 | 1 | N/A |
| U38B | 43.3 | 36.439 | -94.386 | Hyperion 4000 | N4 | 40 | 1 | N/A |
| Y45B* | 536.5 | 33.866 | -89.5431 | Hyperion 4000 | N4 | 40 | 1 | N/A |

### 3.2 Signal Detection and Analysis

The methods for isolating infrasound signals of interest depend on if it was a single-sensor or array station. A brief overview is provided here, but the interested reader is referred to the literature for a more comprehensive review of single sensor (Scott et. al 2007) and array (Le Pichon et al. 2008; Cansi & Le Pichon 2008; Park et al. 2016) analysis. Spectral analysis of single sensor stations can provide waveform and spectral data, but source confirmation requires comparison to ground truth event times and/or other data sources. When single-sensor stations are relatively close to each other and have precise timing, back azimuth and trace velocity can be approximated using trilateration or multilateration (Ruiz et al. 2013; Widdison & Long 2024). This method was utilized with the four single sensor closest stations (121A, 214A, Y22D, and W18A) for the 2016 event (e.g., Figure 4) and all five single sensor stations (U38B, T35B, TUL3, and OSU1) for the 2019 event (see map in Section 4.2.2). In the current work, manual inspection in tandem with a spectrogram detection method (Allen & Rabiner 1977) was used to search for signal arrivals at single stations. The spectrogram detection method utilized a defined noise spectra level to identify regions of increased spectral power within a given frequency band. Continuous wavelet analysis was also performed utilizing the Morlet wavelet (Torrence & Compo 1998) to produce more accurate time of arrival estimates.

In addition to waveform and spectral data, arrays with multiple sensors in close proximity to each other provide direction of arrival (back azimuth), enable high level signal processing, and improve signal-to-noise ratio (SNR). Frequency-wavenumber (*f*-*k*) analysis is a common array method to detect and characterize coherent signals (Blandford 1974; Rost & Thomas 2002). Beamforming algorithms can then compute the signal arrival times, back azimuth, trace velocity, and other relevant parameters of a coherent signal. The open-source software package InfraPy (Blom et al. 2016) was used in the current study to perform beamforming, obtain signal parameters (arrival time, back azimuth, and trace velocity), and evaluate signal signatures (e.g., number of arrivals). The Bartlett beamforming algorithm (Bartlett 1948) with a 10 second time window, 90% overlap, and a frequency band of 0.2 – 3 Hz (frequency band for typical meteor events; Ens et al. 2012) was used for the initial processing.

Further analysis of the signals and yield estimation followed the standard approach for bolide infrasound analysis (Edwards (2009), Ens et al. (2012), Silber 2024b). The frequency band of each detection was the range where the signal power spectral density (PSD) exceeded the noise level prior to detection. Then a bandpass filter corresponding to the detection frequency band was applied to the signal. If there were distinct and separate signal arrivals that fell within the detection window, each detection was analyzed individually. When possible, a 60-second time window was used for the analysis based on previous studies (see Ens et al. 2012); however, the 2019 event duration was too short requiring a reduction to a 10 second window.

The dominant signal period $T$ was obtained using two independent techniques, zero-crossings method (ReVelle 1997) and residual PSD method (Ens et al. 2012). Both are described in detail in Ens et al. (2012), and thus are only briefly mentioned here. The zero-crossings method analyzes the time domain signal at the point of maximum envelope obtained via Hilbert transform (Dziewonski & Hales 1972; Silber et al. 2011). Alternatively, the residual PSD method computes the difference between the signal PSD and the noise PSD. The dominant period is then estimated as the inverse of the frequency with the maximum residual PSD, $T = 1/f_{max}$ (ReVelle 1976).

### 3.3 Infrasound Propagation Modeling

Propagation modeling gives theoretical arrival times that can be used to estimate the shock source, constrain the shock type, and explain the origin of multiple signal arrivals (see Silber & Brown 2014 and Silber 2024a for details). Multiple signal arrivals could be a single signal with multiple propagation paths (e.g., propagation through different atmospheric waveguides) or multiple signals from the same source (e.g., signals from different segments of a meteor trail). InfraGA (Blom 2014), an open-source raytracing package based on geometric acoustics, was used for propagation modeling. Similar codes have been used in other meteor generated infrasound studies (e.g., Silber & Brown 2014; McFadden et al. 2021). InfraGA can run in different coordinate systems (2D/3D Cartesian and spherical; Blom 2019), shoot rays or search for Eigenrays (Blom & Waxler 2017), and implement user-defined atmospheric specifications.

Propagation modeling is directly related to the atmospheric conditions, which are used to calculate the effective speed of sound $\left(c_{eff}\right)$,

$$c_{eff} = \sqrt{\gamma_g R_c T_a} + n_{xy} \cdot \vec{u},$$

where $\gamma_g$ is the ratio of specific heats of air, $R_c$ is the specific gas constant, $T_a$ is the ambient absolute temperature, and $n_{xy} \cdot \vec{u}$ is the wind vector projection in the propagation direction (Gossard and Hooke 1975). Thus, the accuracy of propagation modeling is dependent on the accuracy of the atmospheric profile (i.e., pressure, temperature, density, and zonal/meridional wind speed versus altitude). Here, the National Center for Physical Acoustics hosted the Ground-to-Space (G2S) model (Drob et al. 2003) was used for generating atmospheric profiles. Note that these models cannot capture fine scale fluctuations on the scale of minutes (Averbuch et al. 2022a;

Averbuch et al. 2022b). For the current study, propagation modeling was done using Eigenray search in a spherical coordinate system with range-dependent atmospheric profiles that were updated along the propagation path. G2S atmospheric profiles were obtained for a 60 by 60 evenly spaced grid over the regions of interest. For the 2016 event, the region spanned 30.3° to 41.3° latitude and -121.2° to -103° longitude, while the 2019 event covered 31.8° to 40° latitude and -101.4° to -89.6° longitude. The G2S profiles have a temporal resolution of 1-hour; thus, the closest available time to the event was requested for the two bolides (12:00 UTC for 2016 and 22:00 UTC for 2019).

### 3.4 Energy Estimate

Analysis of the dominant signal period, which is related to the size of the blast cavity produced by the bolide (ReVelle 1976), can be leveraged towards estimating the energy deposition ($E$). For each detection, the dominant period at maximum amplitude was determined using the zero-crossing and PSD methods. If a station received multiple arrivals, for the purpose of calculating energy deposition, only the dominant period associated with the maximum amplitude was used (see Ens et al. 2012). Then the average period across all stations was determined for each method independently. For this average, only the detection containing the maximum Hilbert envelope was considered. Additionally, signals that had apparent thermospheric returns were also excluded (de Groot-Hedlin 2011). Finally, the yield was estimated using an empirical relation that links the energy deposition to the dominant signal period (ReVelle 1997):

$$\log\left(\frac{E}{2}\right) = 3.34\ \log(T) - 2.58 \qquad \frac{E}{2} \leq 100\ \text{kt} \qquad (1)$$

### 3.5 Geostationary Lightning Mapper

The GLM instruments aboard GOES 16 and 17 record images at 500 Hz with a pixel resolution of approximately 8 to 14 km (e.g., Rudlosky et al. 2019). GLM Level 0 (L0) data are digital counts versus time of the measurement, and GLM Level 2 (L2) data are energy (in joules) at time of the event correcting for time of meteoroid flight at an assumed altitude. Lockheed Martin developed calibration tables to convert the L0 and L2 data from digital counts and energy (assuming lightning), respectively, to source intensity assuming blackbody spectral irradiance of 6000 K (Ceplecha et al. 1998). For further information about the utility of GLM in bolide studies, we refer the reader to Smith et al. (2021) and Ozerov et al. (2024). Only the 2019 event was detected by GLM.

# 4 Results and Discussion

## 4.1 June 2016 Bolide

### 4.1.1 June 2016 bolide infrasound detections

The 2016 event was detected at 14 stations, including nine array stations (Table 2). The five single-sensor-stations were part of the USArray Transportable Array (TA) that was funded by the National Science Foundation's EarthScope Initiative (Tytell et al. 2016). Three array stations were IMS infrasound stations (I57US, I58US, and I59US) located in southern California, the Midway Atoll, and Hawaii respectively. Five other array stations were part of the Source Physics Experiment Phase 1 infrasound sensors (IS1-4, IS7) (Snelson et al. 2011; Wilson et al. 2023). Note that array IS6 had a signal that coincided with timing of the event, but the back azimuth was not aligned with this bolide. Finally, one array station in southern California (MR) operated by the University of Hawaii also had a clear detection. Both the TA networks and IMS stations have open data access available from the EarthScope Consortium (previously known as Incorporated Research Institutions for Seismology (IRIS)) database (Trabant et al. 2012; Incorporated Research Institutions for Seismology 2012).

The detections, ranging from 181 km to 6256 km from the estimated USG location (33.8°N, 110.9°W), are listed in Table 3 with signal arrival times and properties. Note that the IS4 and IS7 have nearly a 20° difference between the observed (~145°-146°) and theoretical back azimuth (~127°-128°), which is unusually large but has been previously observed at long range (Silber et al. 2011; Ott et al. 2019; Silber 2024b). A map showing most of the station locations investigated is provided in Figure 3. Omitted stations were due to their location being exceptionally far (I58US and I59US) or co-located with other sensors (TPFO and I57; IS1-7 stations listed as the Nevada National Security Site (NNSS)). Some of the amplitudes (peak-to-peak unless otherwise stated) are listed as not available due to inconsistent and/or unreliable instrument sensitivities. The earliest detection was made at station W18A at 11:03:58 UTC (181 km) and the latest signal was detected at 16:55:47 UTC at station I59US in Midway (6256 km).

Table 3. Detection characteristics at stations for the June 2016 bolide. Peak-to-peak amplitudes ($A_{P2P}$) at stations with unknown sensitivity are listed as N/A. Included are the theoretical back azimuth ($BAZ_T$), observed back azimuth ($BAZ_O$), signal duration ($t_D$), and dominant period ($T$) using the zero-crossing (subscript 'zc') or PSD (subscript 'PSD') method.

| Station | Distance (km) | $BAZ_T$ (deg) | $BAZ_O$ (deg) | Signal Onset (UTC) | $t_D$ (s) | Celerity (km/s) | Frequency Range (Hz) | $A_{P2P}$ (Pa) | $T_{PSD}$ (s) | $T_{zc}$ (s) |
|---|---|---|---|---|---|---|---|---|---|---|
| **W18A** | 181.32 | 216.52 | N/A | 11:04:04 | 15 | 0.401 | 0.125-3 | 2.02 | 2.22 | 2.07 |
| **W18A** | 181.32 | 216.52 | N/A | 11:05:45 | 143 | 0.328 | 0.08-3 | 0.27 | 5.88 | 3.586 |
| **214A** | 271.85 | 40.66 | N/A | 11:12:09 | 71 | 0.290 | 0.1-20 | 0.77 | 5.45 | 5.18 |
| **214A** | 271.85 | 40.66 | N/A | 11:13:20 | 135 | 0.270 | 0.1-20 | 0.81 | 2.73 | 2.896 |
| **214A** | 271.85 | 40.66 | N/A | 11:15:35 | 250 | 0.238 | 0.01-20 | 2.06 | 4.76 | 5.334 |
| **121A** | 322.36 | 296.67 | N/A | 11:13:47 | 106 | 0.311 | 0.1-20 | 0.3 | 4.29 | 2.256 |
| **121A** | 322.36 | 296.67 | N/A | 11:16:27 | 142 | 0.270 | 0.01-20 | 1.92 | 6 | 6.5 |
| **Y22D** | 368.31 | 266.39 | N/A | 11:15:48 | 101 | 0.319 | 0.05-4 | 0.34 | 3 | 4.06 |
| **Y22D** | 368.31 | 266.39 | N/A | 11:18:37 | 207 | 0.278 | 0.035-7 | 0.45 | 5 | 4.53 |
| **TPFO** | 514.17 | 86.07 | N/A | 11:23:00 | 130 | 0.324 | 0.01-7 | 0.16 | 3.16 | 3.43 |
| **TPFO** | 514.17 | 86.07 | N/A | 11:25:10 | 90 | 0.299 | 0.01-7 | 2.34 | 4.6 | 4.396 |
| **TPFO** | 514.17 | 86.07 | N/A | 11:26:40 | 160 | 0.284 | 0.01-7 | 1.41 | 5 | 4.748 |
| **I57** | 514.26 | 86.07 | 83 | 11:23:00 | 130 | 0.324 | 0.01-8 | N/A | 3.16 | 1.45 |
| **I57** | 514.26 | 86.07 | 83 | 11:25:10 | 90 | 0.299 | 0.01-8 | N/A | 4.6 | 4.36 |
| **I57** | 514.26 | 86.07 | 83 | 11:26:40 | 160 | 0.284 | 0.01-8 | N/A | 5 | 1.506 |
| **MR** | 585.12 | 78.39 | 81 | 11:26:33 | 142 | 0.325 | 0.07-3 | 0.25 | 5.45 | 2.526 |
| **MR** | 585.12 | 78.39 | 81 | 11:28:48 | 100 | 0.302 | 0.035-5 | 0.8 | 7.5 | 3.746 |
| **MR** | 585.12 | 78.39 | 81 | 11:30:28 | 254 | 0.287 | 0.01-3 | 3.3 | 3.75 | 4.082 |
| **IS7** | 600.82 | 127.38 | 145.8 | 11:28:29 | 113 | 0.313 | 0.05-10 | N/A | 4.62 | 4.29 |
| **IS7** | 600.82 | 127.38 | 145.8 | 11:30:22 | 101 | 0.296 | 0.02-20 | N/A | 6 | 6.24 |
| **IS7** | 600.82 | 127.38 | 145.8 | 11:32:03 | 240 | 0.282 | 0.05-20 | N/A | 2.86 | 3.02 |
| **IS2** | 602.12 | 127.5 | 128.4 | 11:28:34 | 110 | 0.313 | 0.01-20 | N/A | 2.61 | 4.33 |
| **IS2** | 602.12 | 127.5 | 128.4 | 11:30:24 | 107 | 0.296 | 0.01-20 | N/A | 6 | 6.18 |
| **IS2** | 602.12 | 127.5 | 128.4 | 11:32:11 | 185 | 0.281 | 0.01-20 | N/A | 2.86 | 2.77 |
| **IS4** | 602.18 | 127.55 | 145.2 | 11:28:34 | 115 | 0.313 | 0.05-20 | N/A | 5.45 | 5.53 |
| **IS4** | 602.18 | 127.55 | 145.2 | 11:30:29 | 98 | 0.296 | 0.02-20 | N/A | 6 | 6.19 |
| **IS4** | 602.18 | 127.55 | 145.2 | 11:32:07 | 183 | 0.282 | 0.05-10 | N/A | 2.86 | 2.95 |
| **IS1** | 602.39 | 127.53 | 122.9 | 11:28:34 | 133 | 0.313 | 0.05-20 | N/A | 2.31 | 3.69 |
| **IS1** | 602.39 | 127.53 | 122.9 | 11:30:47 | 82 | 0.293 | 0.02-20 | N/A | 4.62 | 6.21 |
| **IS1** | 602.39 | 127.53 | 122.9 | 11:32:09 | 191 | 0.282 | 0.05-20 | N/A | 3.7 | 2.78 |
| **IS3** | 602.53 | 127.5 | 128 | 11:28:34 | 112 | 0.313 | 0.01-20 | N/A | 3.33 | 4.08 |
| **IS3** | 602.53 | 127.5 | 128 | 11:30:26 | 104 | 0.296 | 0.01-20 | N/A | 4 | 6.19 |
| **IS3** | 602.53 | 127.5 | 128 | 11:32:10 | 190 | 0.282 | 0.05-20 | N/A | 2.86 | 2.75 |
| **I59** | 4698.12 | 61 | 59.56 | 15:15:36 | 590 | 0.302 | 0.04-3 | 0.43 | 4 | 3.96 |
| **I58** | 6256.51 | 66.39 | 74.5 | 16:47:40 | 487 | 0.297 | 0.05-1 | 0.11 | 4 | 2.56 |

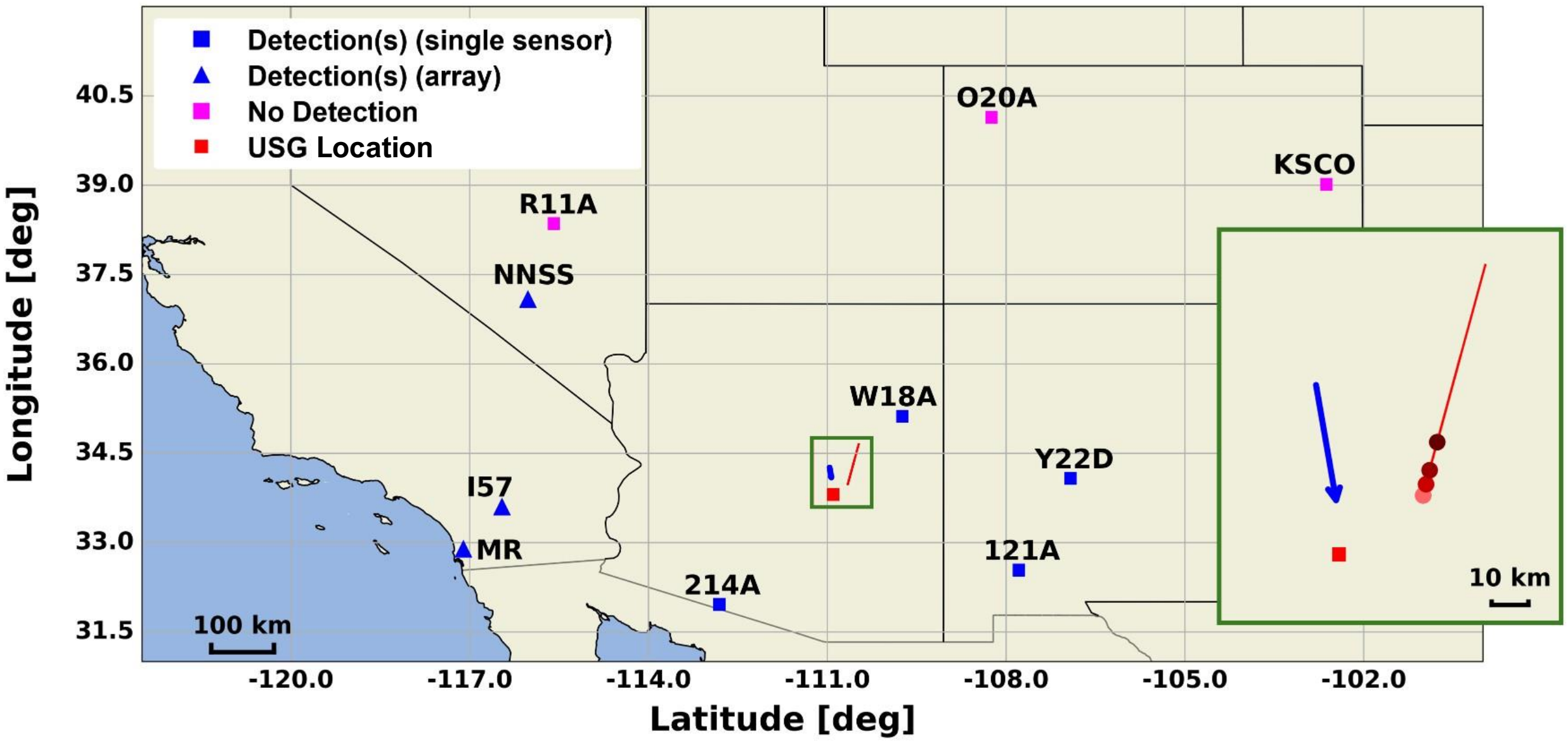

Figure 3. Map of most investigated stations for the 2016 event with or without detections. NNSS denotes the location of six distinct 4-sensor arrays. The inset is a zoomed region near the USG location of peak brightness (red square), which also shows the IMO reported trajectory (blue arrow) and the updated trajectory from Jenniskens et al. (2020) (red line with red circles indicating locations of fragmentation and flare events with lighter color indicating lower altitude).

Figure 4 shows the 2016 bolide signal detections at the stations shown in Figure 3, which each station having two or three distinct detections. This finding is consistent with the video observations and the light curve morphology, indicating that at least some of the “packets” of signals likely originate from distinct fragmentation episodes. While detections were made in all directions, the signal amplitudes were larger for stations to the west (e.g., MR and I57US had amplitudes between 2.5 and 3 Pa while Y22D that was ~180 km closer but to the east had an amplitude of ~0.4 Pa). Consequently, long-range detections favored stations to the west. This directional preference was due to the predominant stratospheric winds blowing westward during this time of the year (summer) in the northern hemisphere at this latitude (Garcés et al. 2004). The average dominant period across all stations using the zero-crossing and PSD methods were 4.77 ± 1.54 s and 4.53 ± 1.15 s, respectively, where the range is the standard deviation over all detections. Of note, two detections (stations 121A and 214A) had an exceptionally strong, low frequency arrival as seen in the blue regions of the respective time traces shown in Figure 4. While the arrival times are consistent with this bolide, the dominant period was too large to be physically related to the bolide energy deposition. Consequently, those dominant periods were omitted in the analysis by recomputing the dominant period for these detections using a 0.1 Hz highpass filter (these results are listed in Table 3).

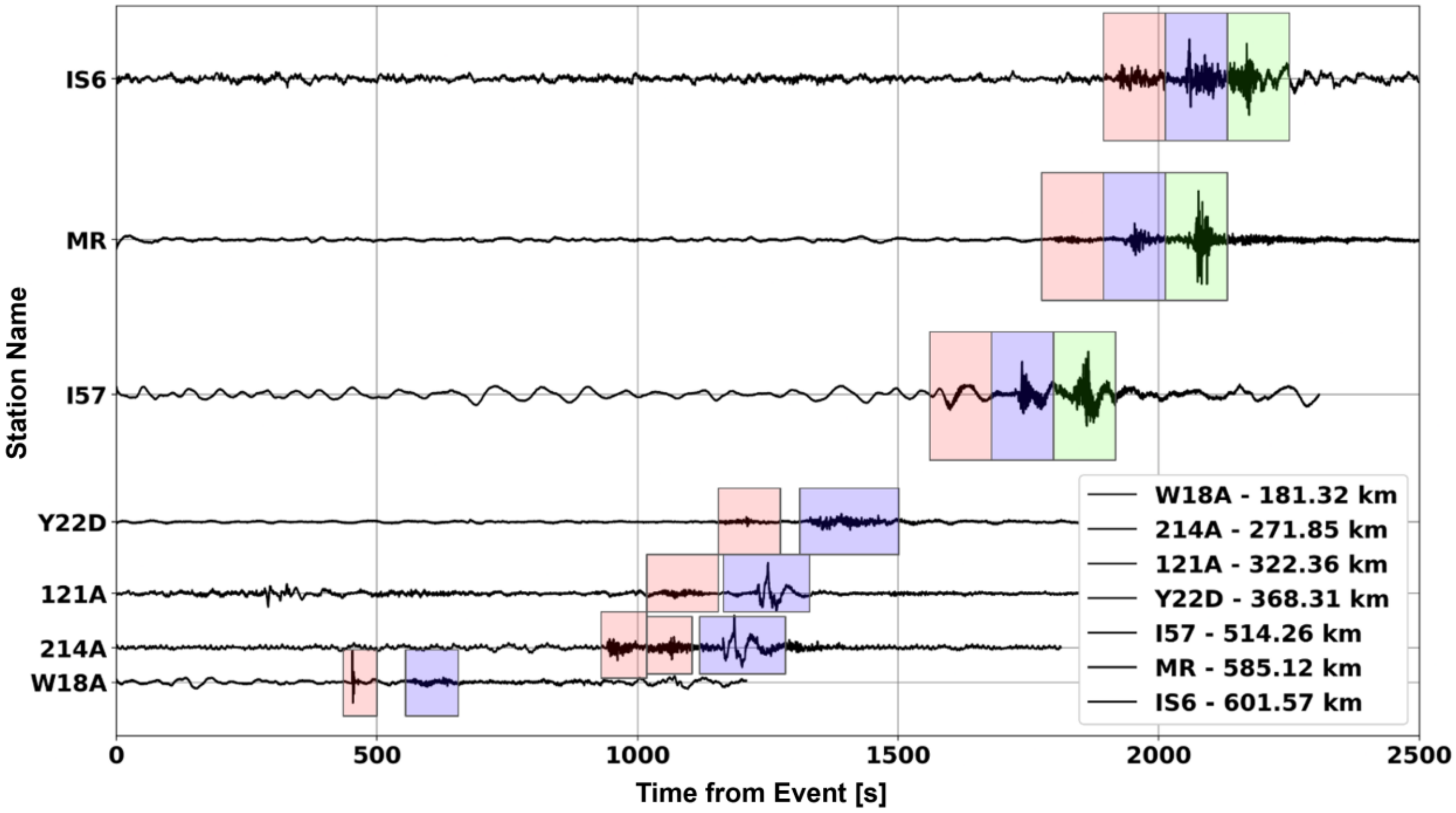

Figure 4: Bandpass filtered (0.01 to 5 Hz) time traces for a subset of stations showing signal arrivals from the June 2016 bolide. Shaded (red, blue, or green) segments indicate periods identified as separate arrivals. The highlighted regions are nominal time windows used for visualization. Additionally, the array station waveforms (I57, MR, IS6) are from a single representative sensor and are not stacked like in the bolide analysis.

Figure 5 shows propagation modeling examples with the source at varying altitudes (based on trajectory information) and station Y22D. Note, the receiver range slightly varies between simulations based on the ground location at different altitudes (Jenniskens et al. 2020). Performing similar simulations for all the stations with detections indicate that most arrivals were likely from the source at altitudes corresponding with fragmentation/flare events, corroborating our earlier assertion about the origin of the observed signals. Only the closest stations indicate possible arrivals from the upper atmosphere. The specific details of these arrivals and the source heights effect on the arriving signal will be discussed later in this section.

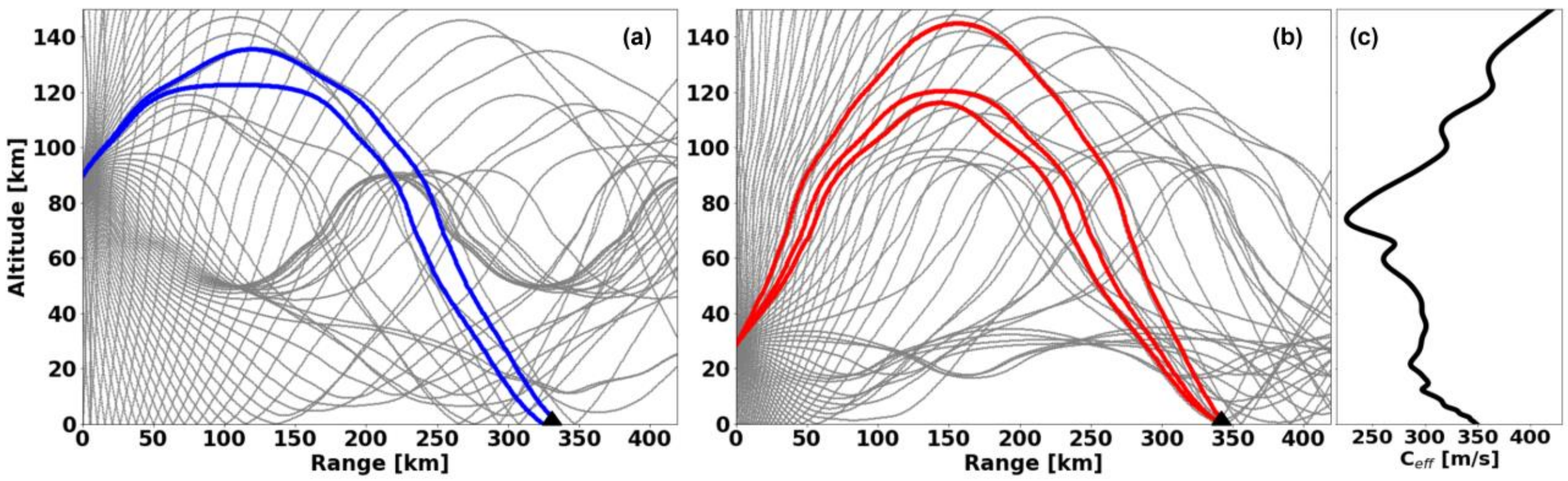

Figure 5. Raytracing simulations for the 2016 event showing propagation paths to station Y22D from a source height of (a) 90 km and (b) 29 km, along with the (c) effective sound speed profile at the source location. Included are the Eigenrays that reach the station (blue or red), Eigenrays that do not reach the station (grey), and the station location (black triangle).

### 4.1.2 Location and trajectory

Infrasound is an independent data source that can be used to further constrain the location and trajectories of bolides. The 2016 event has only a slight difference between the USG reported peak brightness location and ground track information using SACN video analysis (Palotai et al. 2019; Jenniskens et al., 2020). Hence, this event is a good example for assessment of infrasound-based localization. Locations were estimated for single sensor stations within ~400 km based on signal arrival times and celerity, which was obtained through the raytracing simulations. These results are in good agreement with the SACN locations (Jenniskens et al. 2020), which are slightly east of the USG reported location. Additionally, observed back azimuths ($BAZ_o$) from array stations (I57US, MR, and IS1-7) were computed and listed in Table 3 along with the theoretical back azimuths ($BAZ_T$) that is based on the USG reported location. The average difference in BAZ measurements relative to USG location was 4.5°. Analysis relative to the SACN locations (Jenniskens et al. 2020) produce similar results with a difference of 4.1°. Note that IS4 and IS7 had relatively large differences (>15°) and if these are excluded the difference decreases to less than one degree. Thus, the relatively good agreement between both methods demonstrates that infrasound-based localization can produce similar results to other modalities.

### 4.1.3 Cylindrical versus spherical shock differentiation

Analysis of infrasound detections can also assist in assessing if the shock originated from a cylindrical line source or a quasi-spherical source. Cylindrical shocks propagate nearly perpendicular to the trajectory with upward propagation unlikely to reach the ground due to attenuation. Consequently, ground stations typically only detect these shocks when they are lateral and aft of the trajectory at close range (e.g., Silber 2024b). Thus, stations directly forward of the trajectory or at distances beyond ~300 km are unlikely to receive cylindrical shocks from the meteoroid's hypersonic entry (Revelle et al. 2004; Edwards 2009; Silber & Brown 2014; Silber 2024b). This information combined with raytracing suggests that most detections were spherical shocks generated by the fragmentation episodes during the meteoroid's atmospheric entry. Figure 6(a) shows a spherical shock detection from station MR. Almost all other detection, listed in Table 3, were also spherical shock arrivals. However, two stations (W18A, Y22D) had the possibility of cylindrical shock detections.

Station W18A (single sensor) was the most likely to receive a cylindrical shock. Figure 6(b) shows a strong N-wave detection followed by a more diffused signal at W18A. This is classified as a Class I arrival (Silber & Brown 2014), which is characterized by a single N-wave with no other segment having an amplitude more than half the primary for a station within 300 km of the bolide (W18A was ~180 km). The Class I arrival combined with the estimated bolide location (via multilateration) indicates that this arrival is likely the result of a cylindrical shock. This also suggests that W18A is orthogonal to and/or aft of the ground track, which indicates a north to south trajectory. Eigen raytracing indicates that the W18A arrival was likely from an altitude between 50 and 80 km with a ray launched at an angle between 85° and 71°, respectively. This is consistent with past observations (Silber & Brown 2014) that cylindrical shocks typically originate from near the start or center of the trajectory with ray launch angles of 90° ± 25°.

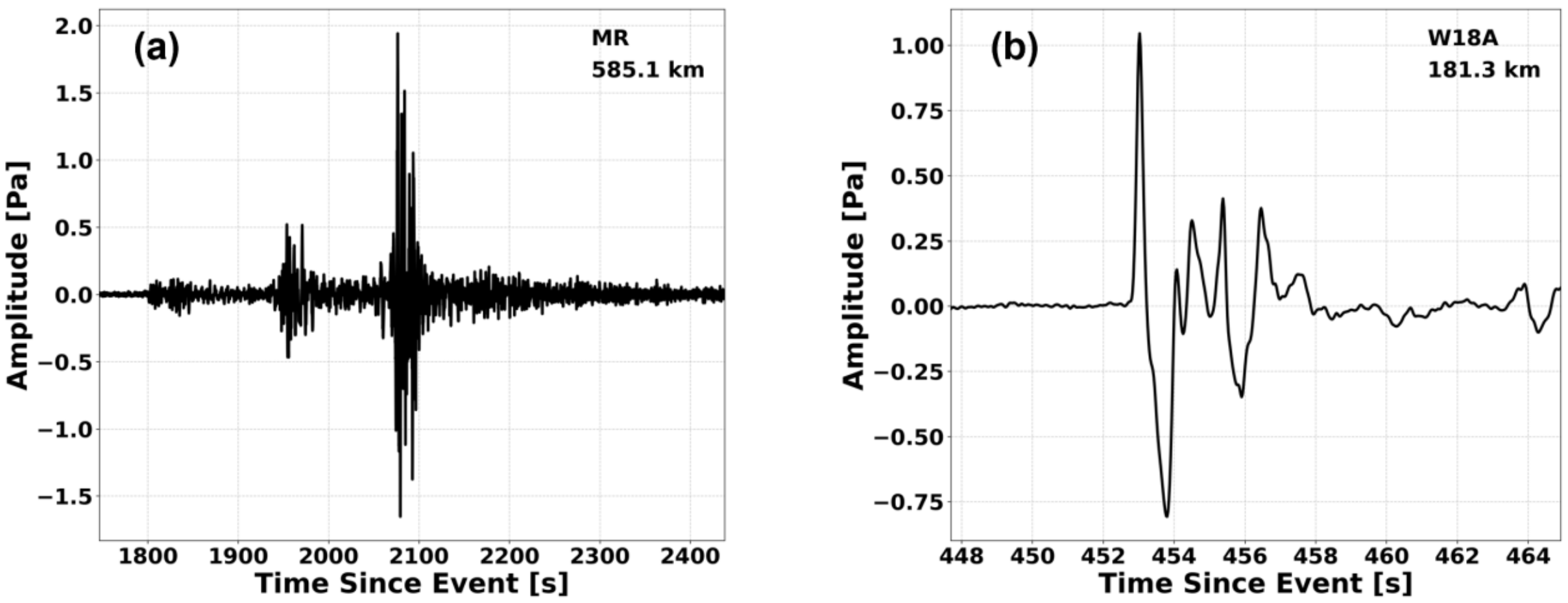

Figure 6. (a) Example of a spherical shock arrival, originating from the bolide's fragmentation events, at station MR. (b) N-wave ballistic shock signal arrival from the 2016 event at station W18A.

Station Y22D had an interesting observation that is supported by raytracing simulations and signal analysis. This station was located just beyond the recommended maximum range for signal classification (Silber & Brown 2014). It received two distinct detections (see Table 3), which Figure 7(a) shows the signal time trace with both detections marked. Figure 7(b) plots the Eigenray arrivals as a function of altitude and time. This shows that Eigenray arrivals at higher altitudes were associated with the first detection. However, the Eigenray arrivals from higher altitudes have an initially upward propagating ray (see Figure 5(a)). This, combined with the fragmentation/flare events being constrained to lower altitudes, suggests that station Y22D received cylindrical shock arrivals that were initially propagating upward. While unlikely due to significant attenuation at higher altitudes, it is possible if the source signal was sufficiently strong with a relatively low ray turning altitude. This could also explain the significantly reduced signal amplitude between the ballistic arrival at W18A (Figure 6(b)) and the first detection at Y22D (Figure 7(a)) because of the increased attenuation at higher altitudes. Similarly, the initially upward propagating cylindrical shock with increased attenuation would explain why the first detection was weaker than the second. Had the initial detection been from a fragmentation event, it would require propagation from a lower altitude, which would have less attenuation. Conversely, the second detection (i.e., red region in Figure 7) has a timing and eigenrays that are aligned with propagation through the lower atmosphere. Since the signal appears non-ballistic in nature with no discernible N-wave pattern, the second detection is likely a spherical shock wave from flare events. Collectively, these results support the ground truth information that estimated a trajectory from northeast to southwest (Palotai et al. 2019) with fragmentation at altitudes ≤ 44 km (Jenniskens et al. 2020). Our study demonstrates that combining infrasound measurements with other sensing modalities, such as optical and space-based observations, improves our ability to characterize bolide events and more clearly link specific signal signatures to their corresponding shock production mechanisms.

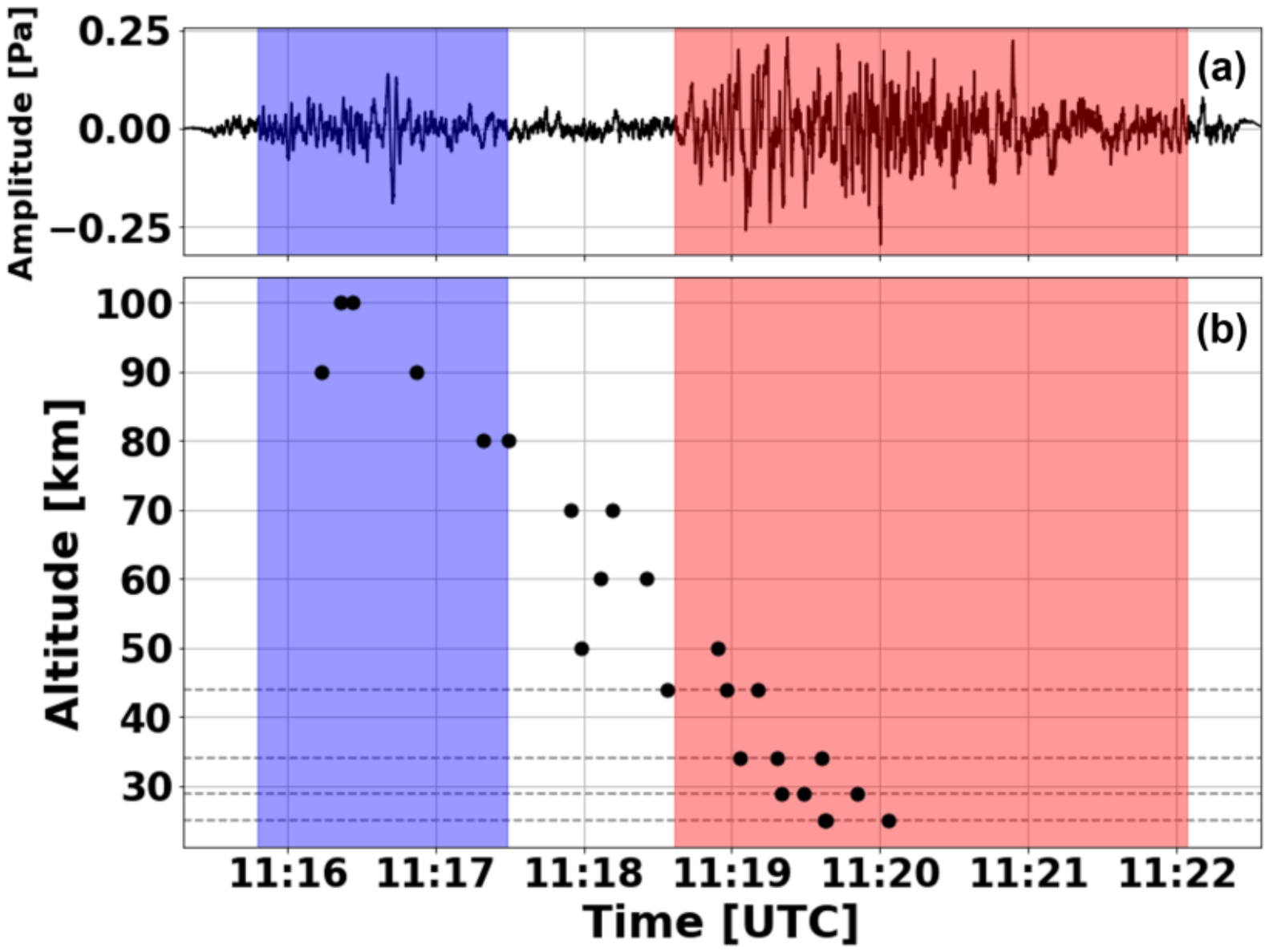

Figure 7. (a) Signal amplitude from station Y22D during the 2019 event with the shaded regions (blue and red) identifying the two distinct detections (see Table 3). (b) Eigenray arrivals at station Y22D for varying source altitudes. Horizontal dashed lines mark ground truth altitudes for fragmentation and flare events (Jenniskens et al. 2020).

### *4.1.4 Yield estimate*

As discussed in Sec. 3.3, empirical relationships can be used to estimate the source energy from the infrasound detections (Silber et al. 2011; Ens et al. 2012; Silber and Brown 2019; Ott et al. 2019). The average dominant period at maximum amplitude using the zero-crossing and PSD methods were 4.77 s and 4.53 s, respectively. Inputting these average periods into Eq. (1) gives an infrasound-based yield estimate of 0.97 kt and 0.82 kt for the zero-crossing and PSD methods, respectively. The average difference between the zero-crossing and PSD methods was 5.2%. These estimates are on the same order magnitude as those reported in the CNEOS database (0.49 kt) and Palotai et al. (2019) (0.54 kt). However, the updated energy estimate from Jenniskens et al. (2020) of 0.047 kt is an order of magnitude less than the other estimates. We note that this apparent discrepancy in energy estimates may suggest a more intricate interplay among bolide physical properties, fragmentation modes, and the altitude at which energy is deposited. Although a comprehensive investigation of these factors is beyond the scope of this paper, addressing them in future studies is essential for a more complete understanding of bolide energy deposition.

## 4.2 April 2019 Bolide

### 4.2.1 GLM detections

The 2019 event was detected on the GLM imagers aboard GOES 16 and 17. However, the GLM-reported locations of peak brightness were further northwest than reported by the CNEOS database. The GLM database also provided an estimated re-navigated altitude of 28 km (none was provided by CNEOS). Our analysis of the GLM pixel locations indicate a north to south trajectory, but there is significant uncertainty since there were only a few data points from GOES 17. The estimated GLM locations are 36.17 N, -94.73 W (GOES 16) and 36.18 N, -94.21 W (GOES 17). Analysis of the GLM light curve utilizing an empirical relationship (Brown et al. 2002) estimated a yield between 0.028 and 0.030 kt, which is slightly lower than the CNEOS based estimate. Given all available information from USG and GLM, it is likely that the 2019 bolide entered the atmosphere near the Oklahoma-Arkansas state line with a north to south heading and experienced two distinct fragmentation events.

### 4.2.2 Infrasound detections

The 2019 event was detected at five stations, including one array station, ranging from 40 to 430 km from the GLM 16 reported location. The array station (OSU1) was a three-sensor infrasound microphone array operated by Oklahoma State University and located in Stillwater, Oklahoma, United States (Wilson et al. 2023). The four single-sensor-stations were part of the N4 network that was operated by the Albuquerque Seismological Laboratory (ASL) and the United States Geological Survey (USGS) (Table 2). These stations were formerly operated within the TA network (TASS Working Group, 2012).

Unlike the 2016 event, the 2019 event lacked multiple distinct arrivals with each detection summarized in Table 4. Figure 8 shows a map of the stations investigated as well as the reported locations of the bolide (USG, GLM aboard GOES 16 and 17). No station east of the bolide had a clear detection, which could be due to a combination of low estimated yield and westward stratospheric winds. However, April is close to the time of year when the seasonal winds in this region change and is therefore more influenced by gravity wave perturbations (Listowski et al. 2024). These factors complicate a definitive explanation for the lack of eastward detections, making it challenging to isolate the contribution of each. The first detection was at 22:22:21 UTC at the closest station (U38B), and the last detection was at 22:46:58 UTC at the farthest station (R32B). Unlike the 2016 event, the amplitude monotonically decreases with increasing distance. The dominant period using the zero-crossing and PSD methods were $0.98 \pm 0.28$ s and $1.00 \pm 0.13$ s, respectively, where the uncertainty is the standard deviation between individual detections. To evaluate the likely source altitudes and propagation paths, range dependent ray tracing simulations were performed, but were unable to constrain the fragmentation events. This will be further discussed in this section.

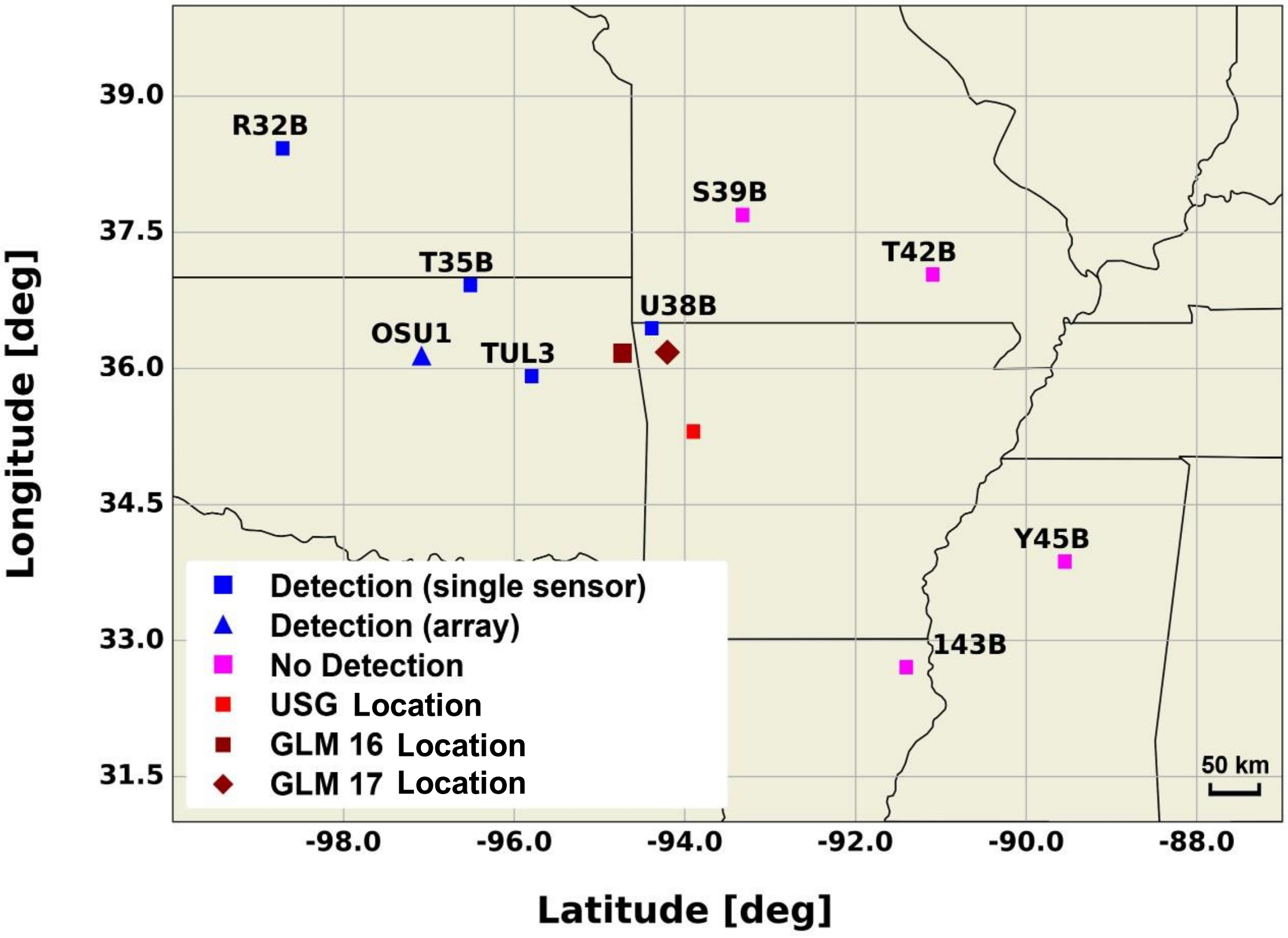

Figure 8. Map of investigated stations for the 2019 event with and without detections as well as the bolide reported location from USG, GLM 16, and GLM 17.

Table 4. Signal detection characteristics at stations for June 2019 bolide. Distances are based on the GLM 16 reported location of maximum brightness. Included are the theoretical back azimuth ($BAZ_T$), observed back azimuth ($BAZ_O$), signal duration $(t_D)$ peak-to-peak amplitude $(A_{P2P})$, and dominant period using the PSD method (subscript 'PSD') and the zero-crossing (subscript 'zc').

| Station | Distance (km) | $BAZ_T$ (deg) | $BAZ_O$ (deg) | Signal Onset (UTC) | $t_D$ (s) | Celerity (km/s) | Frequency Range (Hz) | $A_{P2P}$ (Pa) | $T_{PSD}$ (s) | $T_{zc}$ (s) |
|---|---|---|---|---|---|---|---|---|---|---|
| **U38B** | 43.29 | 225.33 | N/A | 22:22:21 | 12 | 0.216 | 0.2-20 | 7.60 | 0.87 | 0.65 |
| **TUL3** | 99.63 | 73.73 | N/A | 22:25:22 | 19 | 0.261 | 0.05-20 | 4.27 | 1.04 | 0.84 |
| **T35B** | 179.72 | 118.18 | N/A | 22:31:50 | 32 | 0.234 | 0.15-20 | 1.35 | 0.94 | 1.14 |
| **OSU1** | 211.22 | 89.72 | 95 | 22:33:10 | 36 | 0.249 | 0.1-10 | 1.20 | 1.21 | 1.36 |
| **R32B** | 432.33 | 126.66 | N/A | 22:46:18 | 40 | 0.264 | 0.1-9 | 0.51 | 0.95 | 0.91 |

### *4.2.3 Location and trajectory*

The 2019 event has a significant discrepancy (> 100km) between the USG and GLM location estimates. Infrasound provides an independent data source that can further constrain the location and trajectory, which can improve the confidence levels in the location. The one array station (OSU1) with a detection had corresponding back azimuths of ~107° and ~90° for the USG and GLM reported locations, respectively. The observed signal back azimuth was ~95°, which could be combined with the arrival time and celerity to produce a location estimate from this array. However, multilateration is used to leverage the results from the additional detections. Here, circle radii are computed for each detection based on its arrival time and celerity, which was estimated using the ray tracing simulations at varying source altitudes (10-80 km) to match the observed arrival times. The results are shown in Figure 9, where dashed circles around each station are the solutions. The intersection region of the circles shows excellent agreement with the array-based BAZ observation. The infrasound-based location is 30.1 km southeast of GLM 16, 28.2 km southwest of GLM17, and 85.8 km north of USG reports.

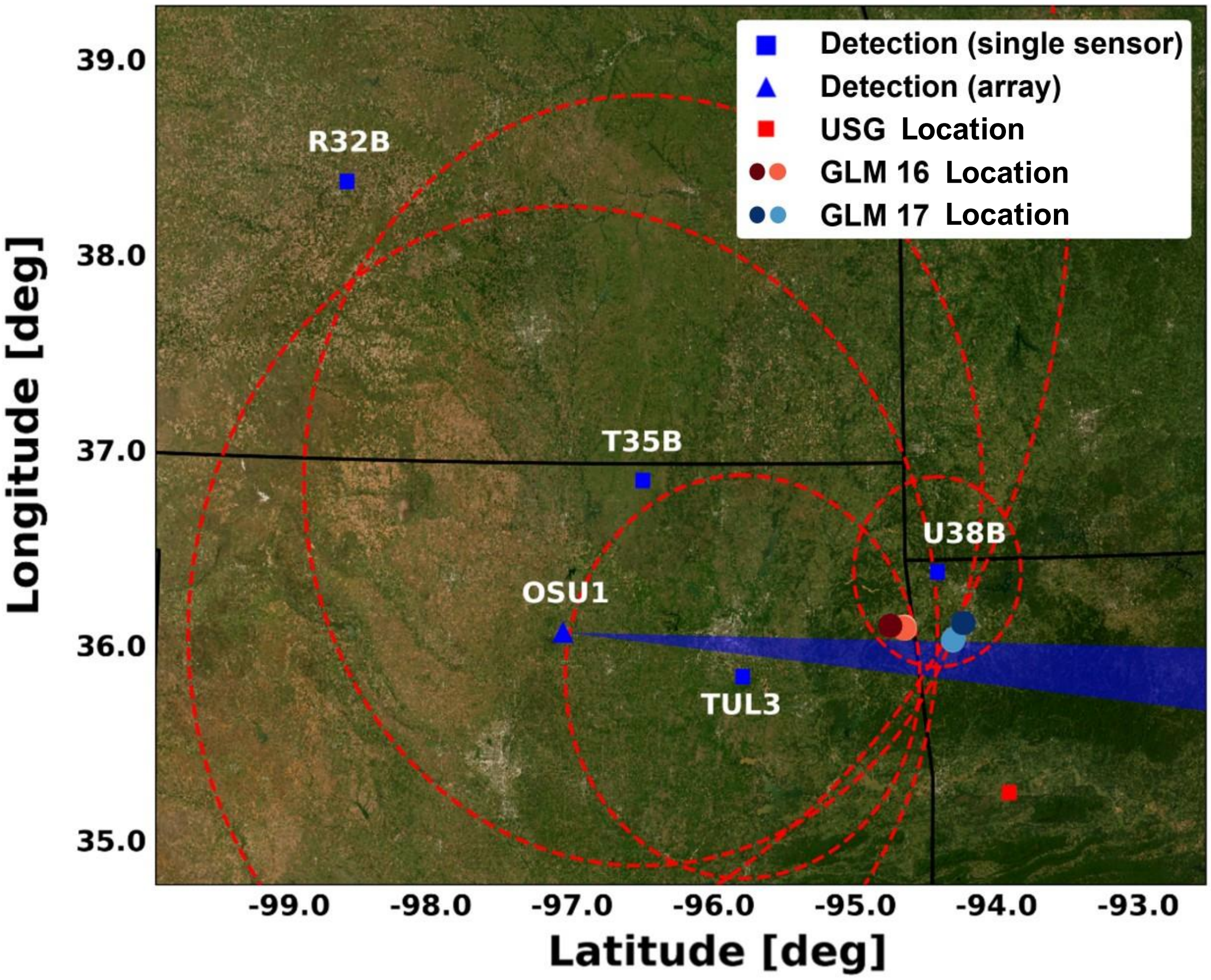

Figure 9. Map showing multilateration (dashed circles centered on each station) and beamforming (blue shaded region) infrasound results for 2019 event along with ground truth information (USG, GLM). The beamforming results (blue shaded region) is the average reported back azimuth with a width of ±1 standard deviation. For the GLM locations, the darker shade is the point of maximum brightness, and the lighter shade is the final reported location.

These infrasound-based detections are more consistent with the GLM location, highlighting how regional-range infrasound detections could constrain event geolocation. However, it is unclear how to appropriately leverage the results from these different modalities. It is expected that a weighted average dependent on the confidence level of each modality would be appropriate. In that respect, Devillepoix (2019) indicated that there are some uncertainties in USG data, notably the reported velocity vector. A recent study (Brown & Borovička 2023) investigated measurement uncertainty for USG sensors by comparing against other reported ground truth. However, the focus of that study was on speed and radiant deviation, for the use case of determining meteor origin. A broader uncertainty analysis for each modality would significantly improve the ability to leverage these independent data sources.

#### *4.2.4 Cylindrical versus spherical shock differentiation*

Like the 2016 bolide, most of the detections are likely from quasi-spherical shocks. Figure 10(a) shows an example of a quasi-spherical shock detection from station OSU1 (Class IV signal as per Silber & Brown (2014) classification scheme). This detection, along with those at R32B and T35B, was determined to be spherical due to their distance from the event and the diffuse nature of the arriving waveform. These three stations are at similar back azimuths and received similar signals with some variation in the signal amplitude. Thus, it is suspected that the three detections propagated from the fragmentation events and traveled in the same atmospheric waveguides. However, two of the stations (U38B, TUL3) were likely cylindrical shock detections. These single sensor detections are shown in Figure 10(b,c). Station U38B was within ~40 km of the bolide and had multiple N-wave arrivals as seen in Figure 10(c). This is classified (Silber & Brown 2014) as a Class II signal with two distinct N-waves. There appears to be a third N-wave (possibly from multipathing), but since its amplitude is less than half the primary it is not Class III. The detection from TUL3, shown in Figure 10(b), was ~100 km from the bolide. The single N-wave arrival followed by a diffuse signal with an amplitude less than half the primary makes this a Class I waveform (Silber & Brown 2014). Eigenray raytracing from altitudes between 10 and 100 km suggest that the source altitude for these detections were 40-50 km and 60-70 km for U38B and TUL3, respectively. Given all available information, these arrivals likely originate from altitudes above those estimated for fragmentation events and the 28 km estimated altitude for the GLM re-navigation.

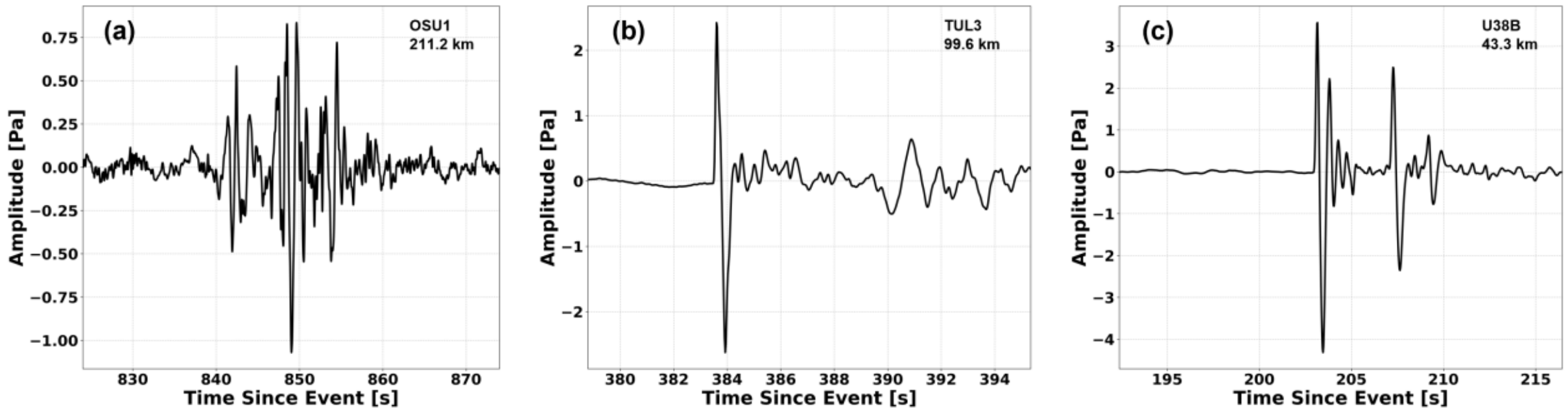

Figure 10. Detections from the 2016 bolide at (a) OSU1, (b) TUL3, and (c) U38B. The diffuse OSU1 detection indicates spherical shock and in contrast to the N-wave ballistic shock arrivals (TUL3, U38B).

#### *4.2.5 Yield estimate*

For the 2019 event, the zero-crossing ($T_{zc} = 0.98$ s) and PSD ($T_{PSD} = 1.00$ s) methods had very similar period estimates. Consequently, the infrasound-based yield was estimated from the average of the two methods using Eq. (1) to be 0.005 kt. This estimate is approximately an order of magnitude smaller than that reported from GLM (0.028 kt) and CNEOS (0.098 kt). While the difference is significant, it is consistent with this event being significantly weaker than the 2016 event and the notably smaller number of detections will increase the uncertainty. As in the 2016 event, this apparent discrepancy might be indicative of other effects that warrant further investigation in future studies.

# 5 Conclusions

In this paper, two relatively low energy bolides were investigated using regional infrasound assets (single sensor and array stations). The first bolide occurred on 2 June 2016 and had many eyewitness accounts and detected on USG sensors. The USG light curve showed two distinct fragmentation events. There were two subsequent studies that leveraged additional observations, including analysis of a recovered sample, to estimate additional meteoroid properties (e.g., trajectory, altitudes, velocity, size, and heading). The second bolide occurred on 4 April 2019 during daylight hours, which limited eyewitness reports. A light curve and a photograph of a smoke trail indicate that it also had two fragmentation episodes. This was a weaker bolide with less ground truth information, which makes it a good example of the potential value of infrasound analysis.

The 2016 bolide had 38 detections at 15 infrasound stations with the farthest detection being over 6000 km from the event. The average dominant period was used with an empirical relationship to produce an infrasound-based energy yield estimate between 0.82 and 0.97 kt. This is consistent with the USG senor and Palotai et al. (2019) analysis, but Jenniskens et al. (2020) reported an order of magnitude lower yield. An infrasound-based location of this bolide was estimated using time of arrival and raytracing for single sensor stations and back azimuths were computed for array stations. These results were in good agreement with the USG sensor reported location. Additional analysis indicated that the majority of detections were spherical shocks generated by fragmentation episodes. However, two of the stations based on the classification of the signal and raytracing suggest a cylindrical shock arrival. Unexpectedly, one of these stations indicates that the cylindrical shock arrival was associated with an initially upward propagating shock.

The 2019 bolide was detected at five stations with only one station having an array. The farthest station with a detection was slightly over 400 km from the event. The dominant period from the infrasound detections produced an estimated yield of 0.005 kt. This yield was significantly lower than that reported by GLM (0.028 kt) and USG (0.098 kt) observations. The cause of the discrepancy is unclear, but this event had significantly fewer detections with a weaker signal (i.e., lower SNR). For this bolide, the ground truth information had significant variation in the reported location (over 100 km difference between USG and GLM sensors), hence infrasound, being an independent data source, can help reduce the location uncertainty. Using multilateration combined with raytracing for single sensor stations as well as the back azimuth of the array station detection, produced an infrasound-based bolide location region that was slightly south of the GLM-based locations. Two of the 2019 detections were likely associated with cylindrical shocks based on signal classification and raytracing. Our findings indicate that in certain cases, well-constrained infrasound observations can help validate or refine USG-reported geolocations, even accounting for inherent uncertainties in both methods.

These two case studies serve as illustrative examples of the potential benefits of combining regional infrasound detections with other established sensing modalities. The 2016 bolide had many detections and more ground truth information that was in good agreement with each other. It served as a demonstration of infrasound being capable of producing estimates that were comparable to these other modalities. For the 2019 bolide being weaker and with limited (and conflicting) ground truth information, regional infrasound detections were able to increase confidence in the event location. This study demonstrates that integrating infrasound data with complementary observations, including optical and space-based measurements, constitutes a rigorous approach to bolide characterization and has the potential to yield a more comprehensive understanding of the underlying shock production mechanisms. In particular, analyzing light curves can help identify the nature of shock production, thereby providing valuable constraints. Furthermore, the findings suggest that infrasound alone, even in the absence of other data, can serve as a viable tool for characterizing high-altitude sources.

## Open Research

Open source code utilized in this study include Los Alamos National Laboratory (LANL) InfraGA infrasound propagation modeling available at https://github.com/LANL-Seismoacoustics/infraGA, and LANL infrasound array processing available at https://github.com/LANL-Seismoacoustics/infrapy. Ground-to-space profiles were obtained from the National Center for Physical Acoustics (NCPA) request site available at https://g2s.ncpa.olemiss.edu. All websites were last accessed in November 2023. The Source Physics Experiments (SPE) would not have been possible without the support of many people from several organizations. The authors wish to express their gratitude to the National Nuclear Security Administration, Defense Nuclear Nonproliferation Research and Development (DNN R&D), and the SPE working group, a multi-institutional and interdisciplinary group of scientists and engineers. This article has been authored by an employee of National Technology & Engineering Solutions of Sandia, LLC under Contract DE-NA0003525 with the U.S. Department of Energy (DOE). 

## Data Availability

The data used in this paper was accessed via the International Federation of Digital Seismograph Networks (FDSN), except for the OS network data. The data used was from the USArray Transportable Array (TA) (IRIS Transportable Array 2003), Central and Eastern US Network (N4) (Albuquerque Seismological Laboratory 2013), International Miscellaneous Station (IM) (Various Institutions 1965), Southern Great Basin Network (SN) (University of Nevada, Reno 1992), UH Infrasound Network (UH) (University of Hawaii), Oklahoma State (OS) (Wilson & Elbing 2023).

### Acknowledgements

The authors thank the anonymous reviewer for their insightful comments that helped refine this paper. This research was supported, in part, by the Gordon and Betty Moore Foundation under grant doi.org/10.37807/gbmf11559. This work was also supported, in part, by the Laboratory Directed Research and Development (LDRD) program at Sandia National Laboratories (project number 229346), a multimission laboratory managed and operated by National Technology and Engineering Solutions of Sandia, LLC., a wholly owned subsidiary of Honeywell International, Inc., for the U.S. Department of Energy's National Nuclear Security Administration under contract DE-NA-0003525. Sandia National Laboratories is a multimission laboratory managed and operated by National Technology and Engineering Solutions of Sandia, LLC (NTESS), a wholly owned subsidiary of Honeywell International Inc., for the U.S. Department of Energy's National Nuclear Security Administration (DOE/NNSA) under contract DE-NA0003525. This written work is authored by an employee of NTESS. The employee, not NTESS, owns the right, title, and interest in and to the written work and is responsible for its contents. Any subjective views or opinions that might be expressed in the written work do not necessarily represent the views of the U.S. Government. The publisher acknowledges that the U.S. Government retains a nonexclusive, paid-up, irrevocable, worldwide license to publish or reproduce the published form of this written work or allow others to do so, for U.S. Government purposes. The DOE will provide public access to results of federally sponsored research in accordance with the DOE Public Access Plan.

## References

Ahn, S.-H., 2003. Meteors and showers a millennium ago. Monthly Notices of the Royal Astronomical Society 343, 1095–1100. https://doi.org/10.1046/j.1365-8711.2003.06752.x

Albuquerque Seismological Laboratory/USGS, 2013. Central and Eastern US Network [dataset]. International Federation of Digital Seismograph Networks. https://doi.org/10.7914/SN/N4

Allen, J.B., Rabiner, L.R., 1977. A unified approach to short-time Fourier analysis and synthesis. Proceedings of the IEEE 65, 1558–1564. https://doi.org/10.1109/PROC.1977.10770

Arrowsmith, S., Negraru, P., Johnson, G., 2021. Bolide Energetics and Infrasound Propagation: Exploring the 18 December 2018 Bering Sea Event to Identify Limitations of Empirical and Numerical Models. The Seismic Record 1, 164–171. https://doi.org/10.1785/0320210034

Assink, J.D., Evers, L.G., Holleman, I., Paulssen, H., 2008. Characterization of infrasound from lightning. Geophysical Research Letters 35. https://doi.org/10.1029/2008GL034193

Averbuch, G., Ronac-Giannone, M., Arrowsmith, S., Anderson, J.F., 2022a. Evidence for Short Temporal Atmospheric Variations Observed by Infrasonic Signals: 1. The Troposphere. Earth and Space Science 9, e2021EA002036. https://doi.org/10.1029/2021EA002036

Averbuch, G., Sabatini, R., Arrowsmith, S., 2022b. Evidence for Short Temporal Atmospheric Variations Observed by Infrasonic Signals: 2. The Stratosphere. Earth and Space Science 9, e2022EA002454. https://doi.org/10.1029/2022EA002454

Bartlett, M.S., 1948. Smoothing Periodograms from Time-Series with Continuous Spectra. Nature 161, 686–687. https://doi.org/10.1038/161686a0

Blandford, R.R., 1974. An automatic event detector at the tonto forest seismic observatory. GEOPHYSICS 39, 633–643. https://doi.org/10.1190/1.1440453

Blom, P., 2019. Modeling infrasonic propagation through a spherical atmospheric layer—Analysis of the stratospheric pair. The Journal of the Acoustical Society of America 145, 2198–2208. https://doi.org/10.1121/1.5096855

Blom, P., 2014. InfraGa/GeoAc: Numerical tools to model acoustic propagation in the geometric limit [WWW Document].

Blom, P., Marcillo, O., Euler, G., 2016. InfraPy: Python-Based Signal Analysis Tools for Infrasound (No. LA-UR--16-24234, 1258366). https://doi.org/10.2172/1258366

Blom, P., Waxler, R., 2017. Modeling and observations of an elevated, moving infrasonic source: Eigenray methods. The Journal of the Acoustical Society of America 141, 2681–2692. https://doi.org/10.1121/1.4980096

Boslough, M., Brown, P., Harris, A., 2015. Updated population and risk assessment for airbursts from near-earth objects (NEOs), in: 2015 IEEE Aerospace Conference. Presented at the 2015 IEEE Aerospace Conference, pp. 1–12. https://doi.org/10.1109/AERO.2015.7119288

Bowman, D.C., Young, E.F., Krishnamoorthy, S., et al., 2018. Geophysical and Planetary Acoustics Via Balloon Borne Platforms (No. SAND2018- 7167R). Sandia National Lab. (SNL-NM), Albuquerque, NM (United States). https://doi.org/10.2172/1459772

Brachet, N., Brown, D., Le Bras, R., et al., 2009. Monitoring the Earth's Atmosphere with the Global IMS Infrasound Network, in: Le Pichon, A., Blanc, E., Hauchecorne, A. (Eds.), Infrasound Monitoring for Atmospheric Studies. Springer Netherlands, Dordrecht, pp. 77–118. https://doi.org/10.1007/978-1-4020-9508-5_3

Brissaud, Q., Krishnamoorthy, S., Jackson, J.M., et al., 2021. The First Detection of an Earthquake From a Balloon Using Its Acoustic Signature. Geophysical Research Letters 48, e2021GL093013. https://doi.org/10.1029/2021GL093013

Brosch, N., Helled, R., Polishook, D., Almoznino, E., David, N., 2004. Meteor light curves: the relevant parameters. Monthly Notices of the Royal Astronomical Society 355, 111–119. https://doi.org/10.1111/j.1365-2966.2004.08300.x

Brown, P., ReVelle, D.O., Silber, E.A., et al., 2008. Analysis of a crater-forming meteorite impact in Peru. Journal of Geophysical Research: Planets 113. https://doi.org/10.1029/2008JE003105

Brown, P., Spalding, R.E., ReVelle, D.O., Tagliaferri, E., Worden, S.P., 2002. The flux of small near-Earth objects colliding with the Earth. Nature 420, 294–296. https://doi.org/10.1038/nature01238

Brown, P.G., Assink, J.D., Astiz, L., Blaauw, R., et al., 2013. A 500-kiloton airburst over Chelyabinsk and an enhanced hazard from small impactors. Nature 503, 238–241. https://doi.org/10.1038/nature12741

Brown, P.G., Borovička, J., 2023. On the Proposed Interstellar Origin of the USG 20140108 Fireball. ApJ 953, 167. https://doi.org/10.3847/1538-4357/ace421

Cansi, Y., Pichon, A.L., 2008. Infrasound Event Detection Using the Progressive Multi-Channel Correlation Algorithm, in: Havelock, D., Kuwano, S., Vorländer, M. (Eds.), Handbook of Signal Processing in Acoustics. Springer, New York, NY, pp. 1425–1435. https://doi.org/10.1007/978-0-387-30441-0_77

Ceplecha, Z., Borovička, J., Elford, W.G., et al., 1998. Meteor Phenomena and Bodies. Space Science Reviews 84, 327–471. https://doi.org/10.1023/A:1005069928850

Chapman, C.R., 2008. Meteoroids, Meteors, and the Near-Earth Object Impact Hazard, in: Trigo-Rodríguez, J.M., Rietmeijer, F.J.M., Llorca, J., Janches, D. (Eds.), Advances in Meteoroid and Meteor Science. Springer, New York, NY, pp. 417–424. https://doi.org/10.1007/978-0-387-78419-9_58

Christie, D.R., Campus, P., 2009. The IMS Infrasound Network: Design and Establishment of Infrasound Stations, in: Le Pichon, A., Blanc, E., Hauchecorne, A. (Eds.), Infrasound Monitoring for Atmospheric Studies. Springer Netherlands, Dordrecht, pp. 29–75. https://doi.org/10.1007/978-1-4020-9508-5_2

Chyba, C.F., Thomas, P.J., Zahnle, K.J., 1993. The 1908 Tunguska explosion: atmospheric disruption of a stony asteroid. Nature 361, 40–44. https://doi.org/10.1038/361040a0

Cobos, M., Antonacci, F., Alexandridis, A., Mouchtaris, A., Lee, B., 2017. A Survey of Sound Source Localization Methods in Wireless Acoustic Sensor Networks. Wireless Communications and Mobile Computing 2017, e3956282. https://doi.org/10.1155/2017/3956282

Cook, R.K., Bedard, A.J., Jr., 1971. On the Measurement of Infrasound. Geophysical Journal International 26, 5–11. https://doi.org/10.1111/j.1365-246X.1971.tb03378.x

C. Plane, J.M., 2012. Cosmic dust in the earth’s atmosphere. Chemical Society Reviews 41, 6507–6518. https://doi.org/10.1039/C2CS35132C

de Groot-Hedlin, C., Hedlin, M.A.H., Walker, K., 2011. Finite difference synthesis of infrasound propagation through a windy, viscous atmosphere: application to a bolide explosion detected by seismic networks. Geophysical Journal International 185, 305–320. https://doi.org/10.1111/j.1365-246X.2010.04925.x

de Groot-Hedlin, C.D., Hedlin, M.A.H., Drob, D.P., 2009. Atmospheric Variability and Infrasound Monitoring, in: Le Pichon, A., Blanc, E., Hauchecorne, A. (Eds.), Infrasound Monitoring for Atmospheric Studies. Springer Netherlands, Dordrecht, pp. 475–507. https://doi.org/10.1007/978-1-4020-9508-5_15

Devillepoix, H.A.R., Bland, P.A., Sansom, E.K., et al., 2019. Observation of metre-scale impactors by the Desert Fireball Network. Monthly Notices of the Royal Astronomical Society 483(4), 5166–5178. https://doi.org/10.1093/mnras/sty3442

Devillepoix, H.A.R., Cupák, M., Bland, P.A., et al., 2020. A Global Fireball Observatory. Planetary and Space Science 191, 105036. https://doi.org/10.1016/j.pss.2020.105036

Drob, D., 2019. Meteorology, Climatology, and Upper Atmospheric Composition for Infrasound Propagation Modeling, in: Le Pichon, A., Blanc, E., Hauchecorne, A. (Eds.), Infrasound Monitoring for Atmospheric Studies: Challenges in Middle Atmosphere Dynamics and Societal Benefits. Springer International Publishing, Cham, pp. 485–508. https://doi.org/10.1007/978-3-319-75140-5_14

Drob, D.P., Garcés, M., Hedlin, M., Brachet, N., 2010. The Temporal Morphology of Infrasound Propagation. Pure Appl. Geophys. 167, 437–453. https://doi.org/10.1007/s00024-010-0080-6

Drob, D.P., Picone, J.M., Garcés, M., 2003. Global morphology of infrasound propagation. Journal of Geophysical Research: Atmospheres 108. https://doi.org/10.1029/2002JD003307

Drolshagen, E., Ott, T., Koschny, D., et al., 2021. Luminous efficiency based on FRIPON meteors and limitations of ablation models. A&A 650, A159. https://doi.org/10.1051/0004-6361/202040204

Dziewonski, A.M., Hales, A.L., 1972. Numerical Analysis of Dispersed Seismic Waves, in: Bolt, B.A. (Ed.), Methods in Computational Physics: Advances in Research and Applications, Seismology: Surface Waves and Earth Oscillations. Elsevier, pp. 39–85. https://doi.org/10.1016/B978-0-12-460811-5.50007-6

Edwards, W.N., 2009. Meteor Generated Infrasound: Theory and Observation, in: Le Pichon, A., Blanc, E., Hauchecorne, A. (Eds.), Infrasound Monitoring for Atmospheric Studies. Springer Netherlands, Dordrecht, pp. 361–414. https://doi.org/10.1007/978-1-4020-9508-5_12

Ens, T.A., Brown, P.G., Edwards, W.N., Silber, E.A., 2012. Infrasound production by bolides: A global statistical study. Journal of Atmospheric and Solar-Terrestrial Physics 80, 208–229. https://doi.org/10.1016/j.jastp.2012.01.018

Fehr, U., 1967. Measurements of infrasound from artificial and natural sources. Journal of Geophysical Research (1896-1977) 72, 2403–2417. https://doi.org/10.1029/JZ072i009p02403

Fernando, B., Mialle, P., Ekström, G., Charalambous, C., Desch, S., Jackson, A., Sansom, E.K., 2024. Seismic and acoustic signals from the 2014 ‘interstellar meteor’, Geophysical Journal International 238(2), 1027–1039. https://doi.org/10.1093/gji/ggae202

Garcés, M., Willis, M., Hetzer, C., Le Pichon, A., Drob, D., 2004. On using ocean swells for continuous infrasonic measurements of winds and temperature in the lower, middle, and upper atmosphere. Geophysical Research Letters 31. https://doi.org/10.1029/2004GL020696

Gi, N., Brown, P., 2017. Refinement of bolide characteristics from infrasound measurements. Planetary and Space Science, SI:Meteoroids 2016 143, 169–181. https://doi.org/10.1016/j.pss.2017.04.021

Gossard, E., Hooke, W., 1975. Waves in the atmosphere : atmospheric infrasound and gravity waves : their generation and propagation.

Gritsevich, M., Koschny, D., 2011. Constraining the luminous efficiency of meteors. Icarus 212, 877–884. https://doi.org/10.1016/j.icarus.2011.01.033

Gritsevich, M.I., 2009. Determination of parameters of meteor bodies based on flight observational data. Advances in Space Research 44, 323–334. https://doi.org/10.1016/j.asr.2009.03.030

Gritsevich, M.I., 2008. Identification of fireball dynamic parameters. Moscow Univ. Mech. Bull. 63, 1–5. https://doi.org/10.1007/s11971-008-1001-5

Gritsevich, M.I., Stulov, V.P., 2008. A model of the motion of the Nneuschwanstein bolide in the atmosphere. Sol Syst Res 42, 118–123. https://doi.org/10.1134/S0038094608020032

Gritsevich, M.I., Stulov, V.P., 2006. Extra-atmospheric masses of the Canadian Network bolides. Sol Syst Res 40, 477–484. https://doi.org/10.1134/S0038094606060050

Harris, A.W., Chodas, P.W., 2021. The population of near-earth asteroids revisited and updated. Icarus 365, 114452. https://doi.org/10.1016/j.icarus.2021.114452

Hawkes, R.L., Bussey, J.E., Macphee, S.L., Pollock, C.S., Taggart, L.W., 2001. Techniques for high resolution meteor light curve investigations 495, 281–286.

Hicks, S.P., Matos, S.B., Pimentel, A., et al., 2023. Exclusive Seismoacoustic Detection and Characterization of an Unseen and Unheard Fireball Over the North Atlantic. Geophysical Research Letters 50, e2023GL105773. https://doi.org/10.1029/2023GL105773

IRIS Transportable Array, 2003. USArray Transportable Array [dataset]. International Federation of Digital Seismograph Networks. https://doi.org/10.7914/SN/TA

Incorporated Research Institutions for Seismology, 2012. Data services products: Infrasound TA infrasound data products.

Jacchia, L.G., 1955. The Physical Theory of Meteors. VIII. Fragmentation as Cause of the Faintmeteor Anomaly. The Astrophysical Journal 121, 521. https://doi.org/10.1086/146012

Jenniskens, P., Albers, J., Tillier, C.E., et al., 2018. Detection of meteoroid impacts by the Geostationary Lightning Mapper on the GOES-16 satellite. Meteoritics & Planetary Science 53, 2445–2469. https://doi.org/10.1111/maps.13137

Jenniskens, P., Moskovitz, N., Garvie, L.A.J., et al., 2020. Orbit and origin of the LL7 chondrite Dishchii'bikoh (Arizona). Meteoritics & Planetary Science 55, 535–557. https://doi.org/10.1111/maps.13452

Kenkmann, T., Artemieva, N.A., Wünnemann, K., et al., 2009. The Carancas meteorite impact crater, Peru: Geologic surveying and modeling of crater formation and atmospheric passage. Meteoritics & Planetary Science 44, 985–1000. https://doi.org/10.1111/j.1945-5100.2009.tb00783.x

Koten, P., Fliegel, K., Vítek, S., Páta, P., 2011. Automatic Video System for Continues Monitoring of the Meteor Activity. Earth Moon Planets 108, 69–76. https://doi.org/10.1007/s11038-011-9380-9

Le Pichon, A., Vergoz, J., Herry, P., Ceranna, L., 2008. Analyzing the detection capability of infrasound arrays in Central Europe. Journal of Geophysical Research: Atmospheres 113. https://doi.org/10.1029/2007JD009509

Listowski, C., Stephan, C.C., Le Pichon, A., et al., 2024. Stratospheric Gravity Waves Impact on Infrasound Transmission Losses Across the International Monitoring System. Pure Appl. Geophys. 1–33. https://doi.org/10.1007/s00024-024-03467-3

Littmann, M., Suomela, T., 2014. Crowdsourcing, the great meteor storm of 1833, and the founding of meteor science. Endeavour 38, 130–138. https://doi.org/10.1016/j.endeavour.2014.03.002

Madiedo, J.M., Trigo-Rodríguez, J.M., 2008. Multi-station Video Orbits of Minor Meteor Showers, in: Trigo-Rodríguez, J. M., Rietmeijer, F.J.M., Llorca, J., Janches, D. (Eds.), Advances in Meteoroid and Meteor Science. Springer, New York, NY, pp. 133–139. https://doi.org/10.1007/978-0-387-78419-9_20

Matoza, R., Fee, D., Green, D., Mialle, P., 2019. Volcano Infrasound and the International Monitoring System, in: Le Pichon, A., Blanc, E., Hauchecorne, A. (Eds.), Infrasound Monitoring for Atmospheric Studies: Challenges in Middle Atmosphere Dynamics and Societal Benefits. Springer International Publishing, Cham, pp. 1023–1077. https://doi.org/10.1007/978-3-319-75140-5_33

McCrosky, R.E., Boeschenstein Jr, H., 1965. The Prairie Meteorite Network. OE 3, 127–132. https://doi.org/10.1117/12.7971304

McFadden, L., Brown, P., Vida, D., Spurný, P., 2021. Fireball characteristics derivable from acoustic data. Journal of Atmospheric and Solar-Terrestrial Physics 216, 105587. https://doi.org/10.1016/j.jastp.2021.105587

Moreno-Ibáñez, M., Silber, E.A., Gritsevich, M., Trigo-Rodríguez, J.M., 2018. Verification of the Flow Regimes Based on High-fidelity Observations of Bright Meteors. ApJ 863, 174. https://doi.org/10.3847/1538-4357/aad334

Moser, D.E., 2017. Comparing eyewitness-derived trajectories of bright meteors to instrumentally-observed data. Planetary and Space Science, SI:Meteoroids 2016 143, 182–191. https://doi.org/10.1016/j.pss.2017.02.016

Negraru, P.T., Golden, P., Herrin, E.T., 2010. Infrasound Propagation in the “Zone of Silence.” Seismological Research Letter 81(4), 259–274.

Nemtchinov, I.V., Svetsov, V.V., Kosarev, I.B., et al., 1997. Assessment of Kinetic Energy of Meteoroids Detected by Satellite-Based Light Sensors. Icarus 130, 259–274. https://doi.org/10.1006/icar.1997.5821

Ott, T., Drolshagen, E., Koschny, D., et al., 2021. Infrasound signals of fireballs detected by the Geostationary Lightning Mapper. A&A 654, A98. https://doi.org/10.1051/0004-6361/202141106

Ott, T., Drolshagen, E., Koschny, D., et al., 2019. Combination of infrasound signals and complementary data for the analysis of bright fireballs. Planetary and Space Science 179, 104715. https://doi.org/10.1016/j.pss.2019.104715

Ozerov, A., Smith, J.C., Dotson, J.L., et al., 2024. GOES GLM, biased bolides, and debiased distributions. Icarus 408, 115843. https://doi.org/10.1016/j.icarus.2023.115843

Palotai, C., Sankar, R., Free, D.L., et al., 2019. Analysis of the 2016 June 2 bolide event over Arizona. Monthly Notices of the Royal Astronomical Society 487, 2307–2318. https://doi.org/10.1093/mnras/stz1424

Park, J., Stump, B.W., Hayward, C., et al., 2016. Detection of regional infrasound signals using array data: Testing, tuning, and physical interpretation. The Journal of the Acoustical Society of America 140, 239–259. https://doi.org/10.1121/1.4954759

Peña-Asensio, E., Trigo-Rodríguez, J.M., Rimola, A., 2022. Orbital Characterization of Superbolides Observed from Space: Dynamical Association with Near-Earth Objects, Meteoroid Streams, and Identification of Hyperbolic Meteoroids. AJ 164, 76. https://doi.org/10.3847/1538-3881/ac75d2

Pichon, A.L., Antier, K., Cansi, Y., et al., 2008. Evidence for a meteoritic origin of the September 15, 2007, Carancas crater. Meteoritics & Planetary Science 43, 1797–1809. https://doi.org/10.1111/j.1945-5100.2008.tb00644.x

Pilger, C., Gaebler, P., Hupe, P., Ott, T., Drolshagen, E., 2020. Global Monitoring and Characterization of Infrasound Signatures by Large Fireballs. Atmosphere 11, 83. https://doi.org/10.3390/atmos11010083

Pilger, C., Hupe, P., Gaebler, P., Ceranna, L., 2021. 1001 Rocket Launches for Space Missions and Their Infrasonic Signature. Geophysical Research Letters 48, e2020GL092262. https://doi.org/10.1029/2020GL092262

Popova, O.P., Jenniskens, P., Emel’yanenko, V., et al., 2013. Chelyabinsk Airburst, Damage Assessment, Meteorite Recovery, and Characterization. Science 342, 1069–1073. https://doi.org/10.1126/science.1242642

Revelle, D.O., 1997. Historical Detection of Atmospheric Impacts by Large Bolides Using Acoustic-Gravity Wavesa. Annals of the New York Academy of Sciences 822, 284–302. https://doi.org/10.1111/j.1749-6632.1997.tb48347.x

ReVelle, D.O., 1976. on meteor-generated infrasound. Journal of Geophysical Research (1896-1977) 81, 1217–1230. https://doi.org/10.1029/JA081i007p01217

Revelle, D.O., Brown, P.G., Spurný, P., 2004. Entry dynamics and acoustics/infrasonic/seismic analysis for the Neuschwanstein meteorite fall. Meteoritics & Planetary Science 39, 1605–1626. https://doi.org/10.1111/j.1945-5100.2004.tb00061.x

ReVelle, D.O., Edwards, W.N., 2007. Stardust—An artificial, low-velocity "meteor" fall and recovery: 15 January 2006. Meteoritics & Planetary Science 42, 271–299. https://doi.org/10.1111/j.1945-5100.2007.tb00232.x

Rost, S., Thomas, C., 2002. Array Seismology: Methods and Applications. Reviews of Geophysics 40, 2-1-2–27. https://doi.org/10.1029/2000RG000100

Rudlosky, S.D., Goodman, S.J., Virts, K.S., Bruning, E.C., 2019. Initial Geostationary Lightning Mapper Observations. Geophysical Research Letters 46, 1097–1104. https://doi.org/10.1029/2018GL081052

Ruiz, D., Ureña, J., García, J.C., et al., 2013. Efficient trilateration algorithm using time differences of arrival. Sensors and Actuators A: Physical 193, 220–232. https://doi.org/10.1016/j.sna.2012.12.021

Sansom, E.K., Gritsevich, M., Devillepoix, H.A.R., et al., 2019. Determining Fireball Fates Using the α–β Criterion. ApJ 885, 115. https://doi.org/10.3847/1538-4357/ab4516

Sansom, E.K., Devillepoix, H.A.R., Yamamoto, M.-y., et al., 2022. The scientific observation campaign of the Hayabusa-2 capsule re-entry, Publications of the Astronomical Society of Japan 74, 50-63. https://doi.org/10.1093/pasj/psab109

Scott, E.D., Hayward, C.T., Kubichek, R.F., et al., 2007. Single and multiple sensor identification of avalanche-generated infrasound. Cold Regions Science and Technology, A Selection of papers presented at the International Snow Science Workshop, Jackson Hole, Wyoming, September 19-24, 2004 47, 159–170. https://doi.org/10.1016/j.coldregions.2006.08.005

Silber, E., Brown, P., 2019. Infrasound Monitoring as a Tool to Characterize Impacting Near-Earth Objects (NEOs), in: Le Pichon, A., Blanc, E., Hauchecorne, A. (Eds.), Infrasound Monitoring for Atmospheric Studies: Challenges in Middle Atmosphere Dynamics and Societal Benefits. Springer International Publishing, Cham, pp. 939–986. https://doi.org/10.1007/978-3-319-75140-5_31

Silber, E.A., 2024a. The Utility of Infrasound in Global Monitoring of Extraterrestrial Impacts: A Case Study of the 2008 July 23 Tajikistan Bolide. AJ 168, 17. https://doi.org/10.3847/1538-3881/ad47c3

Silber, E.A., 2024b. Perspectives and Challenges in Bolide Infrasound Processing and Interpretation: A Focused Review with Case Studies. Remote Sensing 16, 3628. https://doi.org/10.3390/rs16193628

Silber, E.A., Boslough, M., Hocking, W.K., Gritsevich, M., Whitaker, R.W., 2018. Physics of meteor generated shock waves in the Earth's atmosphere – A review. Advances in Space Research 62, 489–532. https://doi.org/10.1016/j.asr.2018.05.010

Silber, E.A., Bowman, D.C., Carr, C.G., et al., 2024. Geophysical Observations of the 24 September 2023 OSIRIS-REx Sample Return Capsule Re-Entry. The Planetary Science Journal 5, 213. https://doi.org/10.3847/PSJ/ad5b5e

Silber, E.A., Bowman, D.C., Ronac Giannone, M., 2023. Detection of the Large Surface Explosion Coupling Experiment by a Sparse Network of Balloon-Borne Infrasound Sensors. Remote Sensing 15, 542. https://doi.org/10.3390/rs15020542

Silber, E.A., Brown, P.G., 2014. Optical observations of meteors generating infrasound—I: Acoustic signal identification and phenomenology. Journal of Atmospheric and Solar-Terrestrial Physics 119, 116–128. https://doi.org/10.1016/j.jastp.2014.07.005

Silber, E.A., Le Pichon, A., Brown, P.G., 2011. Infrasonic detection of a near-Earth object impact over Indonesia on 8 October 2009. Geophysical Research Letters 38. https://doi.org/10.1029/2011GL047633

Silber, E.A., ReVelle, D.O., Brown, P.G., Edwards, W.N., 2009. An estimate of the terrestrial influx of large meteoroids from infrasonic measurements. Journal of Geophysical Research: Planets 114. https://doi.org/10.1029/2009JE003334

Smith, J.C., Morris, R.L., Rumpf, C., et al., 2021. An automated bolide detection pipeline for GOES GLM. Icarus 368, 114576. https://doi.org/10.1016/j.icarus.2021.114576

Snelson, C.M., Barker, D.L., White, R.L., et al., 2011. The Nevada National Security Site - Source Physics Experiment (SPE-N): An Overview.

Subasinghe, D., Campbell-Brown, M.D., Stokan, E., 2016. Physical characteristics of faint meteors by light curve and high-resolution observations, and the implications for parent bodies. Monthly Notices of the Royal Astronomical Society 457, 1289–1298. https://doi.org/10.1093/mnras/stw019

TASS Working Group, 2012. TA Station Selection Strategy and Recommendations  Central and Eastern United States.

Torrence, C., Compo, G.P., 1998. A Practical Guide to Wavelet Analysis. Bulletin of the American Meteorological Society 79, 61–78. https://doi.org/10.1175/1520-0477(1998)079<0061:APGTWA>2.0.CO;2

Trabant, C., Hutko, A.R., Bahavar, M., et al., 2012. Data Products at the IRIS DMC: Stepping Stones for Research and Other Applications. Seismological Research Letters 83, 846–854. https://doi.org/10.1785/0220120032

Trigo-Rodríguez, J.M., Dergham, J., Gritsevich, M., et al., 2021. A Numerical Approach to Study Ablation of Large Bolides: Application to Chelyabinsk. Advances in Astronomy 2021, e8852772. https://doi.org/10.1155/2021/8852772

Tytell, J., Vernon, F., Hedlin, M., et al., 2016. The USArray Transportable Array as a Platform for Weather Observation and Research. Bulletin of the American Meteorological Society 97, 603–619. https://doi.org/10.1175/BAMS-D-14-00204.1

University of Nevada, Reno, 1992. Southern Great Basin Network [dataset]. International Federation of Digital Seismograph Networks. https://doi.org/10.7914/SN/SN

Various Institutions, 1965. International Miscellaneous Stations [dataset]. International Federation of Digital Seismograph Networks. https://doi.org/10.7914/vefq-vh75

Widdison, E., Long, D.G., 2024. A Review of Linear Multilateration Techniques and Applications. IEEE Access 12, 26251–26266. https://doi.org/10.1109/ACCESS.2024.3361835

Wilson, T., Elbing, B. 2023. Infrasonic sources detected on OSU1 [dataset]. Figshare. https://doi.org/10.6084/m9.figshare.22046558.v1

Wilson, T.C., Danneman Dugick, F.K., Bowman, D.C., Petrin, C.E., Elbing, B.R., 2023a. Seismoacoustic Signatures Observed During a Long-Term Deployment of Infrasound Sensors at the Nevada National Security Site. Bulletin of the Seismological Society of America 113, 1493–1512. https://doi.org/10.1785/0120220240

Wilson, T.C., Petrin, C.E., Elbing, B.R., 2023b. Infrasound and Low-Audible Acoustic Detections from a Long-Term Microphone Array Deployment in Oklahoma. Remote Sensing 15, 1455. https://doi.org/10.3390/rs15051455